\documentclass{article}

\usepackage{amsmath}  
\usepackage[english]{babel}
\usepackage[utf8]{inputenc}
\usepackage{amssymb}
\usepackage{graphicx}
\usepackage{authblk}
\usepackage{caption}
\usepackage{multirow}
\usepackage{amscd}
\usepackage{geometry}
\usepackage{a4wide}
\usepackage[usenames]{color}
\usepackage{etaremune} 
\usepackage[hidelinks]{hyperref}
\usepackage{amsfonts}
\usepackage{ifthen}
\usepackage{color}
\usepackage{amscd}
\usepackage{latexsym}

\newcommand{\bi}{\begin{itemize}}
	\newcommand{\ei}{\end{itemize}}
\newcommand{\ben}{\begin{enumerate}}
	\newcommand{\een}{\end{enumerate}}
\newcommand{\beq}{\begin{equation}}
	\newcommand{\eeq}{\end{equation}}
\newcommand{\bd}{\begin{definition}}
	\newcommand{\ed}{\end{definition}}
\newcommand{\bt}{\begin{theorem}}
	\newcommand{\et}{\end{theorem}}
\newcommand{\bcor}{\begin{corollary}}
	\newcommand{\ecor}{\end{corollary}}
\newcommand{\blem}{\begin{lemma}}
	\newcommand{\elem}{\end{lemma}}
\newcommand{\bprop}{\begin{proposition}}
	\newcommand{\eprop}{\end{proposition}}

\newcommand{\R}{\mbox{$ \mathbb{R}  $}}

\newcommand{\h}{\mathcal{H}}
\newcommand{\cc}{\mathbb{C}}

\newcommand{\supp}{\text{supp}}

\newcommand{\bh}{\mathcal{B}(\mathcal{H})}
\newcommand{\bhh}{\mathcal{B}_H(\mathcal{H})}

\def \qed{ \hfill $\Box$ \\}

\newcommand{\bound}{\mathcal{B}}
\newcommand{\Tr}{\text{Tr}}

\newtheorem{definition}{Def.}[section]
\newtheorem{theorem}{Theorem}[section]
\newtheorem{proposition}{Proposition}[section]
\newtheorem{lemma}{Lemma}[section]
\newtheorem{corollary}{Corollary}[section]

\def \proof{\noindent{\it Proof. \;}  \ignorespaces}
\def \qed{ \hfill $\Box$ \\}

\title{On the Monotonicity of relative entropy: A Comparative Study of Petz’s and Uhlmann’s Approaches}

\author[1]{Santiago Matheus\thanks{santiagojavier.matheusrocha@studenti.unipd.it}}
\author[1]{Francesco Bottacin\thanks{bottacin@math.unipd.it}}
\author[2]{Edoardo Provenzi\thanks{edoardo.provenzi@math.u-bordeaux.fr}}
\affil[1]{Università degli studi di Padova, Dipartimento di matematica, Via Trieste, 63, 35121, Padova, Italy}
\affil[2]{Université de Bordeaux, CNRS, Bordeaux INP, IMB, UMR 5251\\ F-33400, 351 Cours de la Libération, Talence, France}

\date{}

\begin{document}

\maketitle

\begin{abstract}
We revisit the monotonicity of relative entropy under the action of quantum channels, a foundational result in quantum information theory. Among the several available proofs, we focus on those by Petz and Uhlmann, which we reformulate within a unified, finite-dimensional operator-theoretic framework. In the first part, we examine Petz’s strategy, identify a subtle flaw in his original use of Jensen’s contractive operator inequality, and point out how it was corrected to restore the validity of his line of reasoning. In the second part, we develop Uhlmann’s approach, which is based on interpolations of positive sesquilinear forms and applies automatically also to non-invertible density operators. By comparing these two approaches, we highlight their complementary strengths: Petz’s method is more direct and clear, Uhlmann’s is more abstract and general. Our treatment aims to clarify the mathematical structure underlying the monotonicity of relative entropy and to make these proofs more accessible to a broader audience interested in both the foundations and the applications of quantum information theory.
\end{abstract}

\section{Introduction}

The (quantum) relative entropy is a central concept in quantum information theory, quantifying the distinguishability between quantum states and serving as a key tool in the analysis of information-processing tasks. One of its most important properties is the \textit{monotonicity under quantum channels\footnote{A quantum channel is a completely positive trace-preserving linear map, see later for a rigorous definition.}}, which expresses the idea that state distinguishability cannot increase during the dynamical evolution of a quantum system that interacts with an environment. This property is also known as Data Processing Inequality (DPI).

The concept of relative entropy was first introduced by Umegaki in 1962 \cite{Umegaki:62}, in the setting of $\sigma$-finite von Neumann algebras, and later extended to arbitrary von Neumann algebras by Araki in 1976 \cite{Araki:76} by means of the Tomita–Takesaki modular theory.

The proof of monotonicity of relative entropy is closely related to the proof of the strong subadditivity (SSA) of the von Neumann entropy. The latter serves as a measure of the mixedness of quantum states, and the validity of the SSA ensures that quantum uncertainty behaves consistently across composite systems, placing fundamental constraints on how information and correlations can be distributed among subsystems.

The first step toward proving SSA was made in 1968 by Lanford and Robinson \cite{lanford1968mean}, who established the subadditivity of the von Neumann entropy and conjectured its stronger form. In 1973, Lieb \cite{lieb1973convex}, building on earlier work by Wigner, Yanase, and Dyson, proved several key properties concerning the convexity and concavity of operator functions and trace functionals. 

These results enabled Lieb and Ruskai to establish the full proof of strong subadditivity of the von Neumann entropy for both finite and infinite-dimensional Hilbert spaces later that same year, in their landmark paper \cite{lieb1973proof}. In that work, they also derived, though without emphasizing it as such, the monotonicity of relative entropy under the partial trace operation, which constitutes a special case of quantum channel.

The first explicit and general proof of the monotonicity of relative entropy under the action of quantum channels was provided two years later by Lindblad in his seminal 1975 paper \cite{lindblad1975completely}, thanks to the results established by Lieb and Ruskai in \cite{lieb1973proof}.

A further breakthrough was achieved by Uhlmann in 1977, who extended the property of monotonicity to a broader class of transformations: the adjoints of unital Schwarz maps, see \cite{uhlmann1977relative}.

The equivalence between SSA and the monotonicity of relative entropy was rigorously established by Petz in the 1980s \cite{Petz:86,Petz:88}. Later, in the work \cite{Petz:2003}, Petz proposed a new proof of monotonicity and posed the question of whether this property also holds for positive (but not necessarily completely positive) trace-preserving linear maps. This question remained open for several years, until it was affirmatively resolved in 2023 by Müller-Hermes and Reeb in \cite{muller-hermes2017monotonicity}. Other pertinent references related to the monotonicity of relative entropy are \cite{sharma:2015,datta:2015,Zhang:2016,Junge:2018}. 

In this paper, we focus specifically on the demonstration strategies proposed by Petz and Uhlmann in the papers \cite{Petz:2003} and \cite{uhlmann1977relative}, respectively, which we believe offer particularly interesting and useful complementary perspectives. Petz’s proof relies on operator-algebraic methods and an explicit representation of the relative entropy in terms of a suitable inner product, an idea inspired by the previously quoted work of Araki.

However, as we show, his original use of Jensen’s operator inequality contains a subtle flaw when applied in its contractive form. We point out how Petz himself and Nielsen corrected the problem in order to restore the validity of Petz’s original approach.

Uhlmann’s proof, on the other hand, is formulated in terms of interpolations of positive sesquilinear forms, a technique that naturally extends to non-invertible states and arbitrary quantum channels. 

Despite their foundational importance, both Petz’s and Uhlmann’s proofs are often considered technically demanding and conceptually opaque. Petz’s approach, while elegant, involves intricate manipulations of operator inequalities that can obscure the overall structure of the argument. Uhlmann’s method introduces a formalism that is unfamiliar to many working in quantum information theory and is rarely presented in full detail in the literature.

We aim to clarify and systematize both strategies by reformulating them in a unified, finite-dimensional operator-theoretic framework, thus making both proofs more accessible to a wider audience.

The paper is structured as follows. In Section 2, we recall the necessary preliminaries on operator convexity, partial trace, and quantum channels. In Section 3, we analyze Petz’s proof, highlight its limitations, and provide a rigorous correction. In Section 4, we develop Uhlmann’s approach in detail and derive the monotonicity of relative entropy in the general setting.

\section{Mathematical preliminaries}
In this section, we start recalling the basic definitions and results needed for the rest of the paper.

Given a finite-dimensional Hilbert state space $(\h,\langle \; ,\; \rangle_\h)$ over the field $\mathbb F=\R$ or $\cc$, $\bh$ indicates the $\mathbb F$-algebra of linear (bounded) operators $A:\h\to \h$. 

We recall that $A\in \bh$ is:
\begin{itemize}
	\item \textit{positive semi-definite}, written $A \ge 0$, if $\langle x,Ax\rangle_\h\ge 0$ for all $x\in \h$;
	\item \textit{positive-definite}, written $A>0$, if $\langle x,Ax\rangle_\h> 0$ for all $x\neq 0_\h\in \h$;
	\item \textit{Hermitian}, if $A=A^\dag$, where $A^\dag$ is the \textit{adjoint} operator of $A$, defined by the formula $\langle A^\dag x,y\rangle_\h = \langle x,Ay\rangle_\h$ for all $x,y\in \h$. 
\end{itemize}
If $\mathbb F=\mathbb C$, then a positive semi-definite, or positive-definite, operator is automatically Hermitian, but this is not the case if $\mathbb F=\mathbb R$.

Suppose now that $\h$ is the state space of a quantum system. The density operator (also called density matrix) $\rho \in \bh$ associated to a given state $s$ of the system is positive semi-definite, Hermitian and such that $\Tr(\rho)=1$. Hence, $\rho$ has eigenvalues $0\le \lambda_j\le 1$ which sum up to 1.

As is well known, if $s$ is a pure state, then $\rho$ is a rank-one orthogonal projector and thus not invertible. 

$\bh$ becomes itself a $\mathbb F$-Hilbert space when it is endowed with the Hilbert-Schmidt operator inner product
\beq
\langle A ,B \rangle_{\bh}:=\Tr(A^\dag B), \quad A,B\in \bh.
\eeq 
The subset of $\bh$ given by Hermitian operators on $\h$, indicated with $\bhh$, is a real Hilbert space with respect to the inner product inherited from $\bh$ and also a partially ordered set with respect to the \textit{Löwner ordering}, defined as follows: for all $A,B\in \bhh$, $A\le B \iff B-A \ge 0$. 

Given a function $f:I\subseteq \R\to \R$ and $A\in \bhh$ with spectral decomposition $A=U DU^\dag$, with $U$ unitary and $D$ diagonal with entries given by the eigenvalues $\lambda_j$ of $A$, all supposed to belong to $I$, we write as usual 
\beq
f(A)=U f(D) U^\dag,
\eeq 
where $f(D)$ is diagonal with non-trivial entries given by $f(\lambda_j)$.

If, for every finite-dimensional $\h$ and every couple of operators $A,B\in \bhh$, we have
\beq 
A \le B \implies f(A) \le f(B),
\eeq 
then $f$ is said to be \textit{operator monotone} on $I$. Instead, if we have 
\beq	
f\left( \frac{A+B}{2} \right) \leq \frac{f(A) + f(B)}{2},
\eeq 
then $f$ is said to be \textit{operator convex} on $I$. If the last inequality is reversed, $f$ is said \textit{operator concave} on $I$.


By Löwner's theorem, see e.g. \cite{Bhatia:1997} chapter V.4, $f:I\to \mathbb{R}$ is operator monotone if and only if it has an analytic continuation that maps the upper half-plane $\mathbb{H}_+$ into itself. Noticeable examples of operator monotone functions on $(0,+\infty)$ are $x\mapsto \log(x)$ and $x\mapsto -1/x$.

If $f:I\to \mathbb{R}$ is a continuous and operator monotone function on $I$, then $\forall a\in I$ the function $F:I\to \mathbb{R}$ given by 
\beq
F(x)=\int_a^x f(t)dt 
\eeq 
is operator convex on $I$, see again \cite{Bhatia:1997}.  So, since $t\mapsto -1/t$ is operator monotone on $(0,+\infty)$, 
\beq
x\longmapsto \int_1^x -\frac{1}{t}dt =-\log(x)
\eeq 
is operator convex on $(0,+\infty)$, which implies that $x\mapsto \log(x)$ is operator concave on $(0,+\infty)$.

As it is well-known, a convex function $f: I \to \mathbb{R}$ satisfies the Jensen inequality
\beq 
f\left( \sum_{i=1}^n \lambda_i x_i \right) \leq \sum_{i=1}^n \lambda_i f(x_i),
\eeq 
for all $x_1, \dots, x_n\in I$ and non-negative weights $\lambda_1, \dots, \lambda_n$ such that $\sum\limits_{i=1}^n \lambda_i = 1$.

The Jensen inequality can be generalized to operator convex functions, as stated in the following celebrated theorem,  proven by Hansen and Pedersen \cite{Hansen:2003}. 

\clearpage

\bt[\textbf{Jensen's operator inequality}]\label{jensen}
	For a continuous function $f:I\to \mathbb{R}$, the following conditions are equivalent.
	\begin{enumerate}
		\item $f$ is operator convex on $I$.
		\item For each natural number $n\geq 1$ the following inequality holds
		\beq \label{operatorjensen}
		f(\sum_{i=1}^n A_i^\dag X_i A_i) \leq \sum_{i=1}^{n} A_i^\dag f(X_i) A_i,
		\eeq
		for every $n$-tuple of bounded Hermitian operators $X_1,\dots,X_n$ on an arbitrary Hilbert space $\h$, with spectra contained in $I$, and every $n$-tuple of operators $(A_1,\dots,A_n)$ on $\h$ satisfying $\sum\limits_{i=1}^n A_i^\dagger A_i= id_\h.$
		\item $f(V^\dag X V)\leq V^\dag f(X) V$ for every isometry $V:\mathcal{K}\to \h$ from an arbitrary Hilbert space $\mathcal{K}$ on an arbitrary Hilbert space $\h$ and every Hermitian operator $X$ on $\h$ with spectrum contained in $I$.
	\end{enumerate}
\et
An immediate consequence of this theorem is the following corollary, also known as the \textit{contractive Jensen's operator inequality}. 
\bcor[\textbf{Contractive version}]
	Let $f:I \to \mathbb{R}$ be a continuous function and suppose that $0\in I$. Then $f$ is operator convex on $I$ and $f(0)\leq 0$ if and only if for some, hence for every, $n\geq 1$ inequality \eqref{operatorjensen} is valid for every n-tuple of bounded Hermitian operators $X_1,\dots,X_n$ on an arbitrary Hilbert space $\h$, with spectra contained in $I$, and every $n$-tuple of operators $(A_1,\dots,A_n)$ on $\h$ satisfying $\sum\limits_{i=1}^n A_i^\dagger A_i\leq id_\h.$
\ecor
The reason for the adjective `contractive' can be easily understood by setting $n=1$: in this case, it follows that $f$ is operator convex on an interval $I$ containing $0$ with $f(0) \le 0$ if and only if
\beq\label{eq:Jensen=1}
f(A^\dagger X A)\leq A^\dagger f(X) A,
\eeq 
for every bounded Hermitian operator $X$ with spectrum in $I$ and every operator $A$ such that $A^\dagger A\leq id$, which implies that $A$ is a contraction, i.e. $\|Ax\|\le \|x\|$ for all $x\in\h$ or, equivalently, $\|A\|\le 1$.

Let us now consider two interacting quantum systems $a$ and $b$ with Hilbert state spaces $\h_a$ and $\h_b$, respectively. The associated composite quantum system $ab$ has Hilbert state space given by $\h_{ab}=\h_a \otimes \h_b$.

In the following, we indicate with $X_a,X_b,X_{ab}$ generic operators of $\mathcal B_H(\h_a)$, $\mathcal B_H(\h_b)$ and $\mathcal B_H(\h_{ab})$, respectively.

The \textit{partial trace} $\Tr_b$ over the subsystem $b$ is a `superoperator', i.e. a linear map $\Tr_b\in \mathcal B(\mathcal B_H(\h_{ab}),\mathcal B_H(\h_a))$, defined as the linear extension to the whole $\mathcal B_H(\h_{ab})$ of the map
\beq\label{eq:proptrace}
\Tr_b(X_a \otimes X_b) = \Tr(X_b)X_a.
\eeq  
$\Tr_b$ satisfies the following property
\beq\label{eq:ptraceb}
\Tr(\Tr_b(X_{ab})X_a) = \Tr(X_{ab}(X_a\otimes id_b)).
\eeq 
see e.g. \cite{Heinosaari:2011}, page 100. 

The partial trace $\Tr_b$ is:
\begin{itemize}
	\item \textit{trace-preserving}: $\Tr(X_{ab})=\Tr(\Tr_b(X_{ab}))$;
	\item \textit{positive}: if $X_{ab} \ge 0$, then $\Tr_b(X_{ab})\ge 0$. 
\end{itemize}
As a consequence, $\Tr_b$ maps states of $\h_{ab}$ into states of $\h_a$. 

Moreover, $\Tr_b$ is \textit{completely positive}, i.e. $\Tr_b\otimes id_\h$ is a positive linear map for all Hilbert spaces $\h$. A trace-preserving and completely positive (or `CPTP') linear map is called a \textit{channel}.

The partial trace is actually one of the three main elements of every channel. In fact, thanks to the Stinespring theorem, see \cite{Stinespring:1955}, given any channel $\mathcal C\in \mathcal B(\mathcal B_H(\h_a),\mathcal B_H(\h_a))$, there exist an operator $Y\in \mathcal B_H(\h_b)$ and a unitary operator $U$ on $\h_a \otimes \h_b$ such that: 
\beq 
\mathcal C(X):= \Tr_b(U(X \otimes Y) U^\dagger), \qquad \forall X\in \bound_H(\h_a).
\eeq 
By the Riesz representation theorem, we can define the adjoint of the partial trace $\Tr_b$ as the only operator $\Tr_b^\dag:\bound_H(\h_a)  \rightarrow  \bound_H(\h_{ab})$ satisfying
\beq\label{eq:T}
\langle X_{ab},\Tr_b^\dag(X_a) \rangle_{\bound_H(\h_{ab})} = \langle \Tr_b(X_{ab}),X_a \rangle_{\bound_H(\h_a)}.
\eeq 
Writing the inner products explicitly and using property \eqref{eq:ptraceb}, we find
\beq
\Tr(X_{ab}\Tr_b^\dag(X_a))=\Tr(\Tr_b(X_{ab})X_a)=\Tr(X_{ab}(X_a\otimes id_b)),
\eeq 
which allows us to write the explicit action of $\Tr_b^\dag$ as
\beq 
\Tr_b^\dag(X_a)=X_a\otimes id_b.
\eeq 
This formula implies that $(\Tr_b^\dag(X_a))^\dag = \Tr_b^\dag(X_a^\dag)$ and that $\Tr_b^\dag$ is \textit{unital}, i.e. it maps the identity of its domain to the identity of its range
\beq
\Tr_b^\dag(id_a)=id_{ab}.
\eeq 
Being a completely positive unital transformation, $\Tr_b^\dag$ is a so-called \textit{Schwarz map}, see \cite{Choi:74} Corollary 2.8, i.e. it satisfies the following inequality:
\beq\label{eq:schwarz}
\Tr_b^\dag(X_a^\dag)  \Tr_b^\dag(X_a) \le \Tr_b^\dag(X_a^\dag \, X_a).
\eeq 
This property is shared by the adjoint of any channel $\cal C$.

\subsection{Relative entropy in quantum information theory}\label{subsec:relentropy}

We devote a separate subsection to relative entropy, as its proper treatment entails the detailed development of several technical aspects. 

This is essential for addressing certain subtleties that are often overlooked in the literature, yet play a crucial role in the rigorous analysis of relative entropy.

Given any finite-dimensional Hilbert space $\h$ and any density operator $\rho\in \bound_H(\h)$ with eigenvalues $\lambda_j\ge 0$, we indicate with $\text{Sp}(\rho)$ its spectrum and with $\text{Sp}_+(\rho)$ the subset of $\text{Sp}(\rho)$ composed only by \textit{strictly positive} eigenvalues. 

It will be convenient to have the following index sets at hand: $I(\rho)=\{j \, : \, \lambda_j \in \text{Sp}(\rho)\}$ and $I_+(\rho)=\{j \, : \, \lambda_j \in \text{Sp}_+(\rho)\}$. 

The \textit{support} of $\rho$, denoted with $\supp(\rho)$, is the subset of $\h$ defined as follows
\beq
\begin{split}
	\supp(\rho)&:=\text{span}\{x_j\in \mathcal H \, : \, x_j \text{ is an  eigenvector of }\rho \text{ with eigenvalue }\lambda_j\in \text{Sp}_+(\rho)\} \\
	&= \bigoplus_{j\in I_+(\rho)} E_{\lambda_j},
\end{split}
\eeq
where $E_{\lambda_j}$ is the eigenspace relative to the eigenvalue $\lambda_j$. 

From this definition and the spectral theorem for Hermitian operators, it immediately follows that
\beq
\supp(\rho)=\ker(\rho)^\perp=\text{Im}(\rho).
\eeq 
This implies that $\h=\ker(\rho)\oplus \supp (\rho)$, so the positive-definite operator $\rho\bigl|_{\supp(\rho)}$ has a trivial kernel, hence it is invertible.

In particular, if $\rho$ is positive-definite then $\rho$ is invertible everywhere and its image and support coincide with the entire $\h$. On the other hand, non-invertible density operators, for instance those corresponding to pure states, have support strictly included in $\mathcal{H}$.

The deviation from purity, or mixedness, of $\rho$ can be measured by its \textit{von Neumann entropy}, defined as follows
\beq
S(\rho):=-\Tr(\rho \log \rho).
\eeq 
The condition $S(\rho)=0$ is satisfied if and only if $\rho$ is pure. In the literature, the precise definition of the logarithm of a matrix may vary according to the specific aims and context in which the matrix logarithm is employed.

For the purposes of this paper, the key property that the logarithm of a density operator $\rho$, or more generally of a Hermitian operator, must satisfy is $\exp(\log\rho) = \rho$. For this reason, we adopt a definition of $\log$ via functional calculus in the extended real numbers with the conventions:
\beq
\log(0)=-\infty, \quad \exp(-\infty)=0, \quad 0 \log 0 = 0,
\eeq 
of course justified by the limits $\log \varepsilon \to -\infty$ as $\varepsilon\to 0^+$, $\exp(-M)\to 0^+$ as $M\to +\infty$, and $\varepsilon \log \varepsilon \to 0$, as $\varepsilon \to 0^+$.

Using the spectral theorem, if $\rho$ is decomposed as
\beq\label{eq:rhospectral}
\rho = \sum_{j\in I_+(\rho)} \lambda_j P_j + 0 P_0,
\eeq 
where $P_j$ denotes the orthogonal projector onto the eigenspace $E_{\lambda_j}$, $j\in I_+(\rho)$, and $P_0$ is the orthogonal projector on $\ker(\rho)$, then, using the previous conventions, we have
\beq
\log \rho := \sum_{j\in I_+(\rho)} \log (\lambda_j) P_j+(-\infty)P_0,
\eeq
and so
\beq 
\exp(\log\rho) = \sum_{j\in I_+(\rho)} \exp(\log \lambda_j) P_j + \exp(-\infty) P_0 = \sum_{j\in I_+(\rho)} \lambda_j P_j +0P_0 = \rho.
\eeq 
Since the projectors satisfy the orthogonality relation $P_i P_j = \delta_{ij} P_j$ and by the convention $0\log 0=0$, which  accounts for the zero eigenvalue, we obtain
\beq
S(\rho)=-\Tr(\rho \log \rho) = -\sum_{j\in I_+(\rho)} \lambda_j \log(\lambda_j) \Tr(P_j),
\eeq 
where $\Tr(P_j)$ is the multiplicity of the eigenvalue $\lambda_j$. In the literature, it is actually more common to write just
\beq
S(\rho) = -\sum_{j \in I_+(\rho)} \lambda_j \log(\lambda_j),
\eeq
with the understanding that each eigenvalue is repeated according to its multiplicity. 

The von Neumann entropy does not, by itself, tell us how one state differs from another. To capture the distinguishability of two states, 
the concept of \textit{relative entropy} (sometimes called the \textit{quantum Kullback–Leibler divergence}) must be introduced.

The definition of the relative entropy between two density operators $\rho$ and $\sigma$ is subjected to a condition on the compatibility of their supports, precisely:
\beq\label{RQEdef}
S(\rho\|\sigma):=\begin{cases}
	\Tr(\rho\log \rho- \rho\log \sigma) & \text{if }\supp(\rho)\subseteq \supp(\sigma) \\
	+\infty & \text{if }\supp(\rho)\nsubseteq \supp(\sigma)
\end{cases},
\eeq
which is known as the `\textit{support-based definition}' of relative entropy. 

In order to understand why the condition $\supp(\rho)\subseteq \supp(\sigma)$ is both necessary and sufficient for $S(\rho\|\sigma)$ to be well-defined, we first notice that 
\beq\label{eq:s2terms}
\Tr(\rho\log \rho- \rho\log \sigma) = \Tr(\rho\log \rho)-\Tr (\rho\log \sigma),
\eeq 
so, the first term of formula \eqref{eq:s2terms} is minus the von Neumann entropy, which is always finite. In order to analyze the second term, let us consider, alongside eq. \eqref{eq:rhospectral}, the analogous spectral decomposition of $\sigma$
\beq
\sigma = \sum_{k\in I_+(\sigma)} \mu_k \Pi_k+0\Pi_0,
\eeq 
so that 
\beq
\log \sigma = \sum_{k\in I_+(\sigma)} \log (\mu_k) \Pi_k+\log(0)\Pi_0,
\eeq
where $\Pi_k$ is the orthogonal projector on the eigenspace relative to the positive eigenvalue $\mu_k$ of $\sigma$, and $\Pi_0$ is the projector on $\ker(\sigma)$. Then, 
\beq\label{equationforcondition}
\begin{split}
	\Tr(\rho \log \sigma) & =\sum_{j\in I_+(\rho)}\sum_{k\in I_+(\sigma)}\lambda_j\log(\mu_k) \Tr({P_j\Pi_k})+\sum_{j\in I_+(\rho)}\lambda_j\log(0)\Tr(P_j\Pi_0),
\end{split}
\eeq 
where the contribution of $P_0$ does not appear due to the convention $0\log(0)=0$. 

The first term is always finite, since only strictly positive eigenvalues appear. Instead, the behavior of the second term depends on the value of $\Tr(P_j\Pi_0)$. To compute it, let us consider any eigenbasis $(x_j)_{j\in I_+(\rho)}$ of $\supp(\rho)$, so that $P_j(x_j)=x_j$ for all $j\in I_+(\rho)$, and use the fact that orthogonal projectors are Hermitian and idempotent to write 
\beq
\begin{split}
	\Tr(P_j\Pi_0) & = \sum_{j\in I_+(\rho)} \langle x_j,P_j\Pi_0 x_j\rangle = \sum_{j\in I_+(\rho)} \langle P_jx_j,\Pi_0 x_j\rangle = \sum_{j\in I_+(\rho)} \langle x_j,\Pi_0 x_j\rangle \\
	& = \sum_{j\in I_+(\rho)} \langle x_j,\Pi_0\Pi_0 x_j\rangle = \sum_{j\in I_+(\rho)} \langle \Pi_0 x_j,\Pi_0 x_j\rangle = \sum_{j\in I_+(\rho)} \|\Pi_0x_j\|^2 \ge 0.
\end{split}
\eeq 
The second term in eq. \eqref{equationforcondition} diverges to $-\infty$ if and only if $\Tr(P_j\Pi_0)>0$, i.e. when there exists at least one $j\in I_+(\rho)$ such that $x_j\in \supp(\rho)$ has a non-trivial projection onto $\ker(\sigma)$, i.e. $\Pi_0x_j$ is non null, which is equivalent to saying that $x_j\notin\ker(\sigma)^\perp=\supp(\sigma)$, therefore 
\beq
\Tr(\rho\log\sigma)=-\infty \iff \Tr(P_j\Pi_0)>0 \iff  \supp(\rho)\nsubseteq \supp(\sigma),
\eeq
which justifies the definition given in \eqref{RQEdef}.

Instead, if $\supp(\rho)\subseteq \supp(\sigma)$, then $\Tr(P_j\Pi_0)=0$ and, using again the convention $0\log(0)=0$, the second term in eq. \eqref{equationforcondition} vanishes and we remain just with the first term, which can be written in an alternative form: consider now the orthonormal eigenbases $(x_j)_{j\in J_+(\rho)}$, $(y_k)_{k\in I_+(\sigma)}$ of $\supp(\rho)$ and $\supp(\sigma)$, respectively, then we have the following well-known formula, see e.g. \cite{Moretti:13},
\beq
\Tr(P_j \Pi_k) = |\langle x_j,y_k\rangle|^2,
\eeq 
which shows that the factor $\Tr(P_j \Pi_k)$ codifies the transition probability between the pure states represented by the unit vectors $x_j$ and $y_k$.

In summary, when $\supp(\rho)\subseteq \supp(\sigma)$, the relative entropy between $\rho$ and $\sigma$ can be written explicitly as follows
\beq\label{quantum_rel_quantum_entropy}
\begin{split}
	S(\rho\|\sigma) & = -S(\rho)-\sum_{j\in I_+(\rho)}\sum_{k\in I_+(\sigma)} \lambda_j \log (\mu_k) \Tr(P_j \Pi_k)\\
	& = \sum_{j \in I_+(\rho)} \lambda_j \left( \log (\lambda_j) - \sum_{k \in I_+(\sigma)} \log (\mu_k) \, |\langle x_j, y_k \rangle|^2 \right).
\end{split}
\eeq 

\clearpage
\noindent Two cases are particularly relevant in practical contexts:
\begin{itemize}
	\item if $\rho,\sigma>0$, then their supports coincide with the entire Hilbert space $\h$ and so their relative entropy is the finite real number given by eq. \eqref{quantum_rel_quantum_entropy};
	\item instead, if $\rho>0$ but $\sigma$ is not, then their relative entropy is infinite. This happens, for instance, when $\rho$ is a full-rank mixed state and $\sigma$ is a pure state.
\end{itemize} 
An equivalent and useful definition of the von Neumann and relative entropy appears in the literature (see, e.g., \cite{wilde2017quantum}). Rather than relying on the support of the density operators, this alternative definition is based on a regularization procedure. 

Specifically, given a state $\rho$ on $\h$, one defines the regularized operator 
\beq
\rho_\varepsilon := \rho + \varepsilon \,id_\h, \qquad \varepsilon > 0.
\eeq 
$\rho_\varepsilon$ is a positive-definite Hermitian operator and the spectral decompositions of $\rho_\varepsilon$ and $\log \rho_\varepsilon$ are
\beq
\rho_\varepsilon=\sum_{j\in I(\rho)}(\lambda_j+\varepsilon)P_j, \quad \log \rho_\varepsilon=\sum_{j\in I(\rho)}\log(\lambda_j+\varepsilon)P_j.
\eeq
We have 
\beq
(\lambda_j+\varepsilon)\log(\lambda_j+\varepsilon) \underset{\varepsilon\to 0^+}{\longrightarrow} \begin{cases}
	\lambda_j \log(\lambda_j)	&   \text{if } j\in I_+(\rho)\\
	0	&   \text{if } \lambda_j=0
\end{cases},
\eeq 
so the von Neumann entropy of $\rho$ is well-defined via the limit
\beq\label{reglimit}
S(\rho):=-\lim_{\varepsilon\to 0^+} \Tr\left(\rho_\varepsilon \log \rho_\varepsilon\right).
\eeq 
Similarly, setting $\sigma_\varepsilon := \sigma + \varepsilon \, id_\h$, $\varepsilon > 0$, the relative entropy between $\rho$ and $\sigma$ can be defined as 
\beq\label{epsilondefofRQE}
S(\rho\|\sigma) :=\lim_{\varepsilon\to 0^+}\Tr( \rho_\varepsilon \log\rho_\varepsilon -\rho_\varepsilon\log \sigma_\varepsilon),
\eeq
known as the `\textit{regularized definition}' of relative entropy.

Let us verify that the regularized and support-based definitions of relative entropy coincide. Equation \eqref{epsilondefofRQE} splits in two terms, the first equals minus the regularized definition of the von Neumann entropy, which is finite by eq. \eqref{reglimit}, so the only issue to address concerns the second term of eq. \eqref{epsilondefofRQE}.
For that, we write the spectral decomposition of $\sigma_\varepsilon$ as follows
\beq
\sigma_\varepsilon=\sum_{k\in I_+(\sigma)}\!(\mu_k+\varepsilon)\Pi_k
+\varepsilon\Pi_0.
\eeq
Repeating the same computations performed in the case of the support-based definition of $S(\rho\|\sigma)$, we obtain
\beq\label{eq:secondterm}
\Tr(\rho_\varepsilon\log\sigma_\varepsilon) 
= \sum_{k\in I_+(\sigma)} \log(\mu_k+\varepsilon)\Tr(\rho_\varepsilon \Pi_k) + \log(\varepsilon)\Tr(\rho_\varepsilon \Pi_0).
\eeq 
If $\supp(\rho)\subseteq\supp(\sigma)$, then 
$\Tr(\rho_\varepsilon\Pi_0)\to 0$ when $\varepsilon\to 0^+$ and the last term in eq. \eqref{eq:secondterm} is absent, so the limit converges to the correct value. 

Instead, if $\supp(\rho)\nsubseteq\supp(\sigma)$, then
$\Tr(\rho_{\varepsilon}\,\Pi_0)\to\alpha> 0$ when $\varepsilon\to 0^+$, consequently the second term in eq. \eqref{eq:secondterm} diverges to $-\infty$, thus matching the behavior of the support-based definition.

The relative entropy has several important properties, see e.g. \cite{Vedral:2002}.
\begin{itemize}
	\item \textit{Klein's inequality}: $S(\rho\|\sigma)\ge 0$ for all $\rho,\sigma$, and $S(\rho\|\sigma)=0$ if and only if $\rho=\sigma$. This property motivates why the relative entropy, despite lacking symmetry in its arguments, is taken to be a measure of distinguishability of states in quantum theories. 
	\item \textit{Invariance under unitary conjugation}: $S(U \rho U^\dagger \| U \sigma U^\dagger) = S(\rho \|  \sigma)$, for all unitary operator $U$ acting on the same Hilbert space as $\rho$ and $\sigma$.
	\item \textit{Additivity w.r.t. tensor product}: $S(\rho_1\otimes\rho_2\|\sigma_1\otimes\sigma_2) = S(\rho_1\|\sigma_1) +S(\rho_2\|\sigma_2)$, for all density operators  $\sigma_j,\rho_j, j=1,2$.
\end{itemize}

\clearpage

\noindent The \textit{monotonicity of $S$ under partial trace} is represented by the inequality
\beq
S(\Tr_b(\rho)\|\Tr_b(\sigma))\leq S(\rho\|\sigma),
\eeq 
which, together with Stinespring’s theorem and the three previously mentioned properties of $S$, permits to prove that quantum distinguishability does not increase under the action of a generic channel $\cal C$
\beq\label{eq:inegrelentropy}
S(\mathcal C(\rho)\|\mathcal C(\sigma))\le S(\rho\|\sigma),
\eeq 
a formula also known as \textit{data processing inequality} (DPI). In fact:
\beq
\begin{split}
	S(\mathcal C(\rho)\|\mathcal C(\sigma))&=S(\Tr_b(U(\rho\otimes X)U^\dagger)\,\|\,\Tr_b(U(\sigma\otimes X)U^\dagger))\\
	&\le S(U(\rho\otimes X)U^\dagger\|U(\sigma \otimes X)U^\dagger)\\
	&=S(\rho\otimes X \|\sigma \otimes X) = S(\rho\|\sigma) + S(X,X)\\
	& =  S(\rho\|\sigma).
\end{split}
\eeq
Inequality \eqref{eq:inegrelentropy} explains why, in the quantum information literature, a channel is often referred to as a \textit{coarse-graining} procedure, a term borrowed from statistical mechanics. 

This terminology reflects the idea that information about different quantum states is lost through the action of the channel, as previously distinguishable states may become indistinguishable after the transformation.

While the first three properties of $S$ mentioned above are relatively straightforward to prove, its monotonicity under partial trace is considerably more subtle. In the next two sections, we provide a detailed analysis of the proof originally proposed by Petz and Uhlmann.

\section{Petz's proof of the monotonicity of the relative entropy under partial trace}\label{sec:Petzproof}

In this section, we examine the strategy proposed by Petz in \cite{Petz:2003} for proving the monotonicity of relative entropy under the partial trace operation, which is based on a clever reformulation of the expression of the relative entropy as a suitable inner product inspired by an analogous construction by Araki \cite{Araki:76}. 

We will show that Petz's proof is flawed due to an incorrect application of the contractive version of Jensen's operator inequality. We will explain how this issue can be circumvented, thus restoring the validity of the overall Petz's approach. Furthermore, we will show how to extend it to incorporate also non-invertible density operators.

The notation that will be used throughout this section is detailed below:
\begin{itemize}
	\item given the finite-dimensional Hilbert spaces $(\h_a,\langle \; ,\; \rangle_a)$ and $(\h_b,\langle \; ,\; \rangle_b)$, we define $\h_{ab}:=\h_a\otimes \h_b$, with inner product $\langle \; ,\; \rangle_{ab}$ induced by those of $\h_a$ and $\h_b$;
	\item operators of $\mathcal B_H(\h_{a})$ will be denoted as $X,Y,Z$ and operators of $\mathcal B_H(\h_{ab})$ by $A,B,C$;
	\item $\rho, \sigma\in\bound_H(\h_{ab})$ are two positive-definite (invertible) density operators\footnote{Actually, for the following analysis, only $\rho$ has to be invertible, however, as we have noted with the support-based definition of the relative entropy, if $\rho>0$, then its support is the entire $\h$ and so, for our analysis to be meaningful, we also have to demand $\sigma >0$.}:  $\rho,\sigma>0$. 
\end{itemize}
In order to rewrite the relative entropy as an inner product, we must introduce some suitable superoperators. 

Precisely, fixed $B,C\in \mathcal B_H(\h_{ab})$, $B>0$, consider the Hermitian superoperators  $R_B,L_C,\Delta_{B,C}\in \mathcal{B}_H(\mathcal B_H(\h_{ab}))$ given by the right and left operator multiplications and the relative modular operator, respectively, i.e.
\beq\label{right,left actions}
R_B(A):=AB, \quad
L_C(A):=C A, \quad
\Delta_{B,C}(A):=L_{C} \circ {R_{B}}^{-1}(A)=C AB^{-1}.
\eeq
Clearly, ${R_B}^{-1}=R_{B^{-1}}$, ${L_C}^{-1}=L_{C^{-1}}$ and 
\beq\label{eq:brackL}
[L_{C},R_{B}]=[L_{C},{R_{B}}^{-1}]=0.
\eeq
All these superoperators are Hermitian, in fact
\beq
\langle R_{B}(A),C \rangle_{ab}  = \Tr((AB)^\dag C) =\Tr(BA C)=\Tr(ACB)=\langle A,R_{B}(C) \rangle_{ab},
\eeq 
and similarly for $L_C$. Regarding $\Delta_{B,C}$, using eq. \eqref{eq:brackL} we have
\beq
\Delta_{B,C}^\dag = (L_{C}{R_{B}}^{-1})^\dag=({R_{B}}^{-1})^\dag L_{C}=(R_{B}^{\dag})^{-1} L_{C}=R_{B}^{-1}\circ L_{C}=L_{C}\circ {R_{B}}^{-1}=\Delta_{B,C}.
\eeq 
If we consider the specific case in which $A=\rho>0$ and $B=\sigma>0$, then we obtain
\beq\label{eq:logdelta}
\begin{split}
	\log(\Delta_{\rho,\sigma}) & = \log(L_{\sigma}\circ R_{\rho}^{-1}) = \log(\exp(\log(L_{\sigma}))\exp(\log(R_{\rho}^{-1}))\\ 
	& = \log(\exp(\log(L_{\sigma})-\log(R_{\rho}))) = \log(L_{\sigma})-\log(R_{\rho}).
\end{split}
\eeq
Since $\log(L_{\sigma})=L_{\log(\sigma)}$ and $\log(R_{\rho})=R_{\log(\rho)}$, applying formula \eqref{eq:logdelta} to $\rho^{1/2}$ gives
\beq
\log(\Delta_{\rho,\sigma})(\rho^{1/2})= \log(\sigma)\rho^{1/2}-\rho^{1/2}\log(\rho).
\eeq 
This last identity, the cyclic property of the trace and the property $[\rho^{1/2},\log(\rho)]=0$ imply
\beq\label{eq:Srhost}
\begin{split}
	S(\rho\|\sigma) & = \Tr[\rho\log(\rho)-\rho\log(\sigma)] = \Tr[\rho^{1/2}(\rho^{1/2}\log(\rho)-\rho^{1/2}\log(\sigma))] \\
	&= \Tr[(\rho^{1/2}\log(\rho)-\rho^{1/2}\log(\sigma))\rho^{1/2}] \\
	& = \Tr[\rho^{1/2}\log(\rho)\rho^{1/2}- \rho^{1/2}\log(\sigma)\rho^{1/2}] \\ 
	& =\Tr[\rho^{1/2}\rho^{1/2}\log(\rho)- \rho^{1/2}\log(\sigma)\rho^{1/2}] \\ 
	& =  -\Tr[\rho^{1/2}\big(\log(\sigma)\rho^{1/2}-\rho^{1/2}\log(\rho)\big)] \\
	& = -\langle \rho^{1/2}, \log(\Delta_{\rho,\sigma}) (\rho^{1/2})\rangle_{ab},
\end{split}
\eeq
which is the `inner product-reformulation' of the relative entropy between the positive-definite states $\rho$ and $\sigma$ that we were searching for. 

We can obtain an analogous formula for the relative entropy of the partial traces of $\rho$ and $\sigma$. To do that, if we fix $Y,Z\in \mathcal B_H(\h_{a})$, $Y>0$, then we can define the Hermitian superoperators $R^a_{Y},L^a_{Z},\Delta^a_{\rho,\sigma}\in \mathcal{B}_H(\mathcal B_H(\h_{a}))$ as follows
\beq
R^a_{Y}(X):=XY, \quad 
L^a_{Z}(X):=ZX, \quad
\Delta^a_{Y,Z}(X):=L_Z\circ {R_{Y}}^{-1}(X)=ZXY^{-1}.
\eeq
By carrying out computations analogous to those in eq. \eqref{eq:Srhost}, but this time using the superoperators $R^a_{\Tr_b(\rho)}, L^a_{\Tr_b(\sigma)},\Delta^a_{\rho,\sigma}:=L^a_{\Tr_b(\sigma)} \circ R^a_{\Tr_b(\rho)^{-1}}$, we obtain
\beq\label{eq:Srhost2}
S(\Tr_b(\rho)\| \Tr_b(\sigma))=-\langle \Tr_b(\rho)^{1/2}, \log(\Delta^a_{\rho,\sigma})\, (\Tr_b(\rho)^{1/2})\rangle_{a} \, .
\eeq
Due to the minus sign in front of the inner products appearing in eqs. \eqref{eq:Srhost} and \eqref{eq:Srhost2}, we have that \textit{the monotonicity of the relative entropy under partial trace is equivalent to}:
\beq\label{eq:reformulation}
\langle \rho^{1/2}, \log(\Delta_{\rho,\sigma} )(\rho^{1/2})\rangle_{ab}  \le \langle \Tr_b(\rho)^{1/2}, \log(\Delta^a_{\rho,\sigma})(\Tr_b(\rho)^{1/2})\rangle_{a} \, .
\eeq
These inner products are defined on two different Hilbert spaces, $\bound_H(\h_{ab})$ and $\bound_H(\h_a)$, in order to perform a meaningful comparison and prove the inequality, Petz introduced a superoperator $V_\rho:\bound_H(\h_a)\to \bound_H(\h_{ab})$ through the explicit formula
\beq\label{defVs}
V_\rho(X\,\Tr_b(\rho)^{1/2}) := \Tr_b^\dag(X)\,\rho^{1/2},
\eeq
which serves as a bridge between the reduced state $\operatorname{Tr}_b(\rho)$ and the full state $\rho$. While, as we are going to show, this definition is computationally effective, it may at first appear somewhat ad hoc. Actually, a seemingly more natural choice for $V_\rho$ would have been $\Tr_b^\dag$ for two reasons: first, as for $V_\rho$, $\Tr_b^\dag$ is a map between $\bound_H(\h_a)$ and $\bound_H(\h_{ab})$ and, second, it satisfies the Schwarz inequality \eqref{eq:schwarz}, which in the following will play a crucial role in the proof of inequality \eqref{eq:reformulation}.

It turns out that $V_\rho$ is tightly related to the adjoint of the partial trace $\Tr_b$, not w.r.t. the original inner products of $\h_a$ and $\h_{ab}$, but w.r.t. suitably \emph{weighted inner products} that naturally emerge from the previous structural analysis of the relative entropy in terms of superoperators.

In fact, the reformulations of the relative entropy obtained in eqs. \eqref{eq:Srhost} and \eqref{eq:Srhost2}, respectively, involve inner products weighted by $\rho^{1/2}$ and $\Tr_b(\rho)^{1/2}$. This observation suggests that the correct notion of adjoint to consider for $\Tr_b$ is the one defined relative to the inner products\footnote{The positive-definiteness of these inner products is guaranteed by the fact that $\rho,\Tr_b(\rho)>0$.}
\beq
\langle A,B \rangle_{ab,\, \rho} := \langle R_{\rho^{-1/2}}(A) , B \rangle_{ab}, \quad	\langle X,Y \rangle_{a,\, \rho} :=  \langle R^a_{\Tr_b(\rho)^{-1/2}}(X) ,Y \rangle_{a}.
\eeq
We have
\beq
\begin{split}
	\langle X,\Tr_b(A)\rangle_{a,\, \rho}&= \langle R^a_{\Tr_b(\rho)^{-1/2}}(X) ,\Tr_b(A) \rangle_{a} = \langle \Tr_b^\dag \circ {R^a_{\Tr_b(\rho)^{-1/2}}}(X),A\rangle_{ab}  \\
	& =  \langle R_{\rho^{-1/2}} \circ R_{\rho^{1/2}} \circ \Tr_b^\dag \circ {R^a_{\Tr_b(\rho)^{-1/2}}}(X),A\rangle_{ab} \\
	& = \langle R_{\rho^{1/2}} \circ \Tr_b^\dag \circ {R^a_{\Tr_b(\rho)^{-1/2}}}(X),A\rangle_{ab,\, \rho}.
\end{split}
\eeq
So, the adjoint of $\Tr_b$ w.r.t. the weighted inner products defined above is the operator $\Tr_b^{\dag,\, \rho}:=R_{\rho^{1/2}} \circ \Tr_b^\dag \circ R^a_{\Tr_b(\rho)^{-1/2}}$, which, for all $X\in \bound_H(\h_a)$, satisfies
\beq
(R_{\rho^{1/2}} \circ \Tr_b^\dag \circ R^a_{\Tr_b(\rho)^{-1/2}})(X\, \Tr_b(\rho)^{1/2}) = (R_{\rho^{1/2}}\circ \Tr_b^\dag)(X)=\Tr_b^\dag(X)\, \rho^{1/2},
\eeq 
therefore $V_\rho=\Tr_b^{\dag,\, \rho}$, i.e. 
\beq\label{eq:defVs}
V_\rho = R_{\rho^{1/2}} \circ \Tr_b^\dag \circ R^a_{\Tr_b(\rho)^{-1/2}}.
\eeq
From this fact, we obtain that the explicit action of $V_\rho$ on any $X\in \bound_H(\h_a)$ is 
\beq\label{eq:VsonX}
V_\rho(X) = (X\,\Tr_b(\rho)^{-1/2} \otimes id_b) \, \rho^{1/2},
\eeq
and this implies immediately that $V_\rho$ transforms $\Tr_b(\rho)^{1/2}$ into $\rho^{1/2}$
\beq\label{Vsidentity}
V_\rho(\Tr_b(\rho)^{1/2})=\rho^{1/2}.
\eeq
Using repeatedly the cyclic property of the trace and  Schwarz's inequality \eqref{eq:schwarz} satisfied by $\Tr_b^\dag$, we can prove that $V_\rho$ is a contraction
\beq\label{vscontraction}
\begin{split} 
	\|V_\rho(X \Tr_b(\rho)^{1/2})\|^2& = \langle \Tr_b^\dag(X) \rho^{1/2},\Tr_b^\dag(X) \rho^{1/2}\rangle_{ab} = \Tr[(\Tr_b^\dag(X)\rho^{1/2})^\dag \Tr_b^\dag(X) \rho^{1/2}]\\
	&  = \Tr[\rho^{1/2} \Tr_b^\dag(X^\dag) \Tr_b^\dag(X) \rho^{1/2}] = \Tr[\Tr_b^\dag(X^\dag) \Tr_b^\dag(X) \rho] \\
	& \le \Tr[\Tr_b^\dag(X^\dag X) \rho]=\langle \Tr_b^\dag(X^\dag X),\rho\rangle_{ab}=\langle X^\dag X,\Tr_b(\rho)\rangle_a\\
	& = \Tr(X^\dag X \Tr_b(\rho)^{1/2} \Tr_b(\rho)^{1/2})= \Tr[\Tr_b(\rho)^{1/2}X^\dag X\Tr_b(\rho)^{1/2}]\\
	& = \Tr[(X\Tr_b(\rho)^{1/2})^\dag X \Tr_b(\rho)^{1/2}]=\langle X \Tr_b(\rho)^{1/2},X \Tr_b(\rho)^{1/2}\rangle_{a}\\
	& =\|X \Tr_b(\rho)^{1/2}\|^2.
\end{split}
\eeq 
Note now that
\beq\label{eq:DeltaaTr}
\begin{split}
	\langle \Delta^a_{\rho,\sigma} (X)\Tr_b(\rho)^{1/2}, X\Tr_b(\rho)^{1/2}\rangle_a & 
	= \langle \Tr_b(\sigma)X \Tr_b(\rho)^{-1/2}, X\Tr_b(\rho)^{1/2}\rangle_a\\
	& =  \Tr[(\Tr_b(\sigma)X \Tr_b(\rho)^{-1/2})^\dag X\Tr_b(\rho)^{1/2}] \\
	&  = \Tr[\Tr_b(\rho)^{-1/2}X^\dag\Tr_b(\sigma) X\Tr_b(\rho)^{1/2}] \\ 
	& = \Tr[XX^\dagger \Tr_b(\sigma)].
\end{split}
\eeq
Using the equality just proven, we can show that $V_\rho^\dag \Delta_{\rho,\sigma} V_\rho\leq \Delta^a_{\rho,\sigma}$, in fact
\beq
\begin{split}
	\langle V_\rho^\dag \Delta_{\rho,\sigma} V_\rho (X\Tr_b(\rho)^{1/2}), X \Tr_b(\rho)^{1/2}\rangle_a &= \langle \Delta_{\rho,\sigma} V_\rho (X\Tr_b(\rho)^{1/2}), V_\rho(X \Tr_b(\rho)^{1/2})\rangle_{ab}\\
	&=\langle \Delta_{\rho,\sigma} (\Tr_b^\dag(X) \rho^{1/2}), \Tr_b^\dag(X) \rho^{1/2}\rangle_{ab}	\\
	& =\langle \sigma\Tr_b^\dag(X) \rho^{-1/2}, \Tr_b^\dag(X)\rho^{1/2}\rangle\\
	& = \Tr[\rho^{-1/2}\Tr_b^\dag(X^\dagger)\sigma\Tr_b^\dag(X) \rho^{1/2}] \\
	& = \Tr[\Tr_b^\dag(X) \Tr_b^\dag(X^\dagger)\sigma] \\
	& \le \Tr[\Tr_b^\dag(XX^\dagger)\sigma ]= \langle \Tr_b^\dag(XX^\dagger),\sigma\rangle_{ab} \\
	& = \langle XX^\dagger,\Tr_b(\sigma)\rangle_{a} = \text{Tr}[XX^\dagger  \Tr_b(\sigma)]\\
	& =  \langle \Delta^a_{\rho,\sigma} (X)\Tr_b(\rho)^{1/2}, X\Tr_b(\rho)^{1/2}\rangle_a,
\end{split}
\label{vsdeltavs<delta0}
\eeq
where we have used again Schwarz's inequality and, to write the last equality, we have applied eq. \eqref{eq:DeltaaTr}.

Since $\log(x) $ is an operator monotone function, the inequality $V_\rho^\dag \Delta_{\rho,\sigma} V_\rho\leq \Delta^a_{\rho,\sigma}$ implies
\begin{equation}
	\log(V_\rho^\dag \Delta_{\rho,\sigma} V_\rho)\leq \log(\Delta^a_{\rho,\sigma}),
	\label{monotoneineq}
\end{equation}
and so
\beq
\begin{split}
	S(\Tr_b(\rho)\|\Tr_b(\sigma))& =\langle \Tr_b(\rho)^{1/2}, -\log(\Delta^a_{\rho,\sigma}) (\Tr_b(\rho)^{1/2})\rangle_a \\
	& \le \langle \Tr_b(\rho)^{1/2}, -\log(V_\rho^\dag \Delta_{\rho,\sigma} V_\rho)(\Tr_b(\rho)^{1/2})\rangle_a.
\end{split}
\eeq
Petz's strategy to make the relative entropy $S(\rho\|\sigma)$ appear on the right-hand side of the previous inequality consists in using the fact that $V_\rho$ is a contraction to apply the contractive Jensen operator inequality \eqref{eq:Jensen=1} with $f(x)=-\log(x)$. In this way, due to \eqref{Vsidentity}, we would get
\beq\label{eq:pseudoproof}
\begin{split}
	S(\Tr_b(\rho)\|\Tr_b(\sigma)) & \le \langle \Tr_b(\rho)^{1/2}, -\log(V_\rho^\dag \Delta_{\rho,\sigma} V_\rho)(\Tr_b(\rho)^{1/2})\rangle_a \\
	& \le \langle \Tr_b(\rho)^{1/2}, V_\rho^\dag(-\log( \Delta_{\rho,\sigma}))V_\rho(\Tr_b(\rho)^{1/2})\rangle_a \\
	& = \langle V_\rho\Tr_b(\rho)^{1/2},-\log( \Delta_{\rho,\sigma})V_\rho(\Tr_b(\rho)^{1/2})\rangle_{ab}\\
	&=\langle \rho^{1/2}, -\log(\Delta_{\rho,\sigma}) (\rho^{1/2})\rangle_{ab}\\
	&=S(\rho\|\sigma),
\end{split}
\eeq
and so the proof of the monotonicity of $S$ w.r.t. the partial trace $\Tr_b$ would be achieved.

For this argument to be valid, $-\log(x)$ is required to be operator convex, which is true, to be well defined at $x=0$, and to satisfy $-\log(0)\le 0$. 

Petz circumvented the lack of definition of $-\log(x)$ in $x=0$ by using the following integral identity
\begin{equation}\label{transforming the log to 1/x}
	\begin{split}
		\int_0^{+\infty} \left(\frac{1}{x+\xi}-\frac{1}{1+\xi}\right) d\xi&=\lim_{M\to +\infty} \left. \big(\log(x+\xi)-\log(1+\xi)\big)\right\vert_{0}^{M}=\\
		&=\lim_{M\to +\infty }\log\left(\frac{x+M}{1+M}\right)-\log(x)\\ & = -\log(x).
	\end{split}
\end{equation}
Since the integral over $[0,+\infty)$ coincides with that over $(0,+\infty)$, we may restrict our attention to strictly positive values of $\xi$. If we denote by $id_{ab}$ the identity operator on $\bound_H(\h_{ab})$, we have
\beq
\Delta_{\rho,\sigma,\xi}\equiv(\Delta_{\rho,\sigma}+\xi id_{ab})^{-1} -(id_{ab}+\xi id_{ab})^{-1} = (\Delta_{\rho,\sigma}+\xi id_{ab})^{-1} -id_{ab}(1+\xi)^{-1}, 
\eeq
moreover
\beq\label{eq:traceone}
\langle \rho^{1/2}, -id_{ab}(1+\xi)^{-1} \rho^{1/2}\rangle_{ab} = - (1+\xi)^{-1} \Tr(\rho)= -(1+\xi)^{-1},
\eeq 
thus
\beq
\langle \rho^{1/2},\Delta_{s,t,\xi}\, \rho^{1/2}\rangle_{ab} = \langle \rho^{1/2}, (\Delta_{\rho,\sigma}+\xi id_{ab})^{-1}\rho^{1/2}\rangle_{ab} - (1+\xi)^{-1}.
\eeq
It follows that
\beq
S(\rho\| \sigma)= \int_0^{\infty} \left( \langle \rho^{1/2}, (\Delta_{\rho,\sigma}+\xi id_{ab})^{-1}\rho^{1/2}\rangle_{ab} - (1+\xi)^{-1} \right) d\xi,
\eeq 
and, analogously,
\beq    
S(\Tr_b(\rho)\| \Tr_b(\sigma))= \int_0^{\infty} \left(\langle \Tr_b(\rho)^{1/2}, (\Delta_{\rho,\sigma}^a+\xi id_{a})^{-1} \Tr_b(\rho)^{1/2}\rangle_a-(1+\xi)^{-1}\right) d\xi.
\eeq
For all $\xi\in (0,+\infty)$, the function $g_\xi$ given by $x\mapsto (x+\xi)^{-1}-(1+\xi)^{-1}$ is well defined for $x=0$, it is operator convex and operator monotone (decreasing), see \cite{Bhatia:1997} chapter V.1. Thanks to this last property we have
\beq \label{monotone-dec-ineq}
(\Delta^a_{\rho,\sigma}+\xi)^{-1}\leq (V_\rho^\dag \Delta^a_{\rho,\sigma} V_\rho+\xi)^{-1}.
\eeq
However, $g_\xi(0)=\xi^{-1}-(1+\xi)^{-1}>0$ for all $\xi\in(0,+\infty)$, so \textit{the contractive Jensen operator inequality cannot be used to write} \beq\label{eq:fails}
(V_\rho^\dag \Delta^a_{\rho,\sigma} V_\rho+\xi)^{-1}\leq V_\rho^\dag (\Delta^a_{\rho,\sigma}+\xi)^{-1} V_\rho,
\eeq 
which would lead to the proof of the monotonicity of the relative entropy with computation analogous to those shown in formula  \eqref{eq:pseudoproof}. 

\begin{figure}[!ht]
	\centering
	\includegraphics[width=0.6\textwidth]{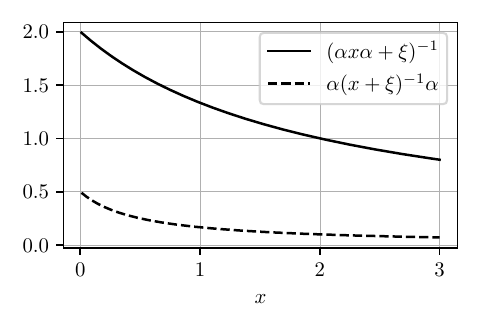}
	\caption{Comparison of $(\alpha x\alpha+\xi)^{-1}$ and $\alpha(x+\xi)^{-1}\alpha$, illustrating the failure of the Jensen-type inequality in the scalar case, with $\alpha=0.5$ and $\xi=0.5$.}
	\label{1/(x+t)}
\end{figure}

A simple counterexample illustrating the failure of inequality \eqref{eq:fails} arises in the scalar case, where a contraction reduces to a multiplication by a real coefficient $\alpha \in (0,1]$, and satisfies $\alpha^\dag = \alpha$. The inequality
\beq\label{eq:false}
(\alpha x \alpha + \xi)^{-1} \leq \alpha(x+\xi)^{-1} \alpha, 
\eeq
is \textit{false} for all $\xi \in (0,+\infty)$, as shown in Figure \ref{1/(x+t)}.

\begin{figure}[!ht]
	\centering
	\includegraphics[width=0.6\textwidth]{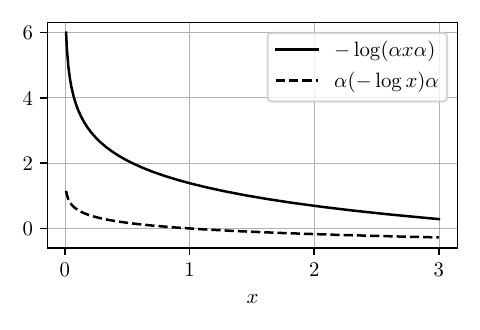}
	\caption{Counterexample showing that the inequality $-\log(\alpha x \alpha) \leq -\alpha \log(x)\alpha$ fails for $\alpha=0.5$.}
	\label{-log}
\end{figure}

Note that the inequality $-\log(V^\dagger X V)\leq -V^\dagger\log(X)V$ is also false for a generic contraction $V$, as we show in Figure \ref{-log} with a counterexample in the scalar case.

To summarize, showing that $V_\rho$ is a contraction does not permit to use the contractive version of Jensen's operator inequality to prove the monotonicity of the relative entropy. 

\subsection{Correction of Petz's strategy to prove the relative entropy monotonicity}\label{susbsec:Petzcorrection}

There is a simple correction of Petz's strategy that restores the validity of his approach to prove the monotonicity of the relative entropy in a rigorous way. The correction was provided by Petz himself, together with Nielsen, in \cite{NielsenPetz:2005}. The same line of reasoning can be found also in \cite{Tomamichel:2009} and \cite{KhatriWilde:2024}.

The key idea of the correction lies in recognizing that the operator $V_\rho$
is not merely a contraction, but an isometry, i.e. $V_\rho^\dag V_\rho$ is the identity operator on $\bound_H(\h_{a})$. In fact, using eqs. \eqref{eq:ptraceb} and \eqref{defVs}, we have
\begin{equation}\label{Vsisometry}
	\begin{split}
		\langle Y\Tr_b(\rho)^{1/2}, V^\dagger_\rho V_\rho(X\Tr_b(\rho)^{1/2})\rangle_a &=\langle V_\rho(Y\Tr_b(\rho)^{1/2}), V_\rho(X\Tr_b(\rho)^{1/2})\rangle_{ab}\\
		&= \langle \Tr_b^\dag(Y) \rho^{1/2}, \Tr_b^\dag(X)\rho^{1/2}\rangle_{ab}\\
		&=\Tr[(Y^\dagger\otimes id_b)(X\otimes id_b)\rho]\\
		&=\Tr[(Y^\dagger X\otimes id_b) \rho]=\Tr[Y^\dagger X \Tr_b(\rho)]\\
		&=\Tr[\Tr_b(\rho)^{1/2} Y^\dagger X \Tr_b(\rho)^{1/2}]\\
		&=\Tr[(Y\Tr_b(\rho)^{1/2})^\dagger X \Tr_b(\rho)^{1/2}]\\
		&=\langle Y \Tr_b(\rho)^{1/2}, X \Tr_b(\rho)^{1/2}\rangle_{a}.
	\end{split}
\end{equation}
This allows us to apply point $(iii)$ of Theorem \ref{jensen} with $\h=\bound_H(\h_{ab})$, $\mathcal{K}=\bound_H(\h_a)$ and $X=\Delta_{\rho,\sigma}$, which ensures that the inequality
\beq
-\log(V^\dagger_s \Delta_{\rho,\sigma} V_\rho)\leq -V^\dagger_s\log(\Delta_{\rho,\sigma})V_\rho,
\eeq
holds true, confirming the validity of the computations in \eqref{eq:pseudoproof}, thereby establishing the monotonicity of the relative entropy.

\subsection{Extension of Petz's proof to non-invertible density operators}

The corrected version of Petz’s strategy for proving the monotonicity of relative entropy can be extended to include non-invertible density operators. To this end, the interplay between the support-based and the regularized definition of relative entropy given in \eqref{RQEdef} and \eqref{epsilondefofRQE}, respectively, will prove to be useful. 

First of all, note that if $\supp(\rho)\nsubseteq\supp(\sigma)$, then $S(\rho\|\sigma) = +\infty$, and the monotonicity statement is trivially true. 

We therefore restrict our attention to the case $\supp(\rho) \subseteq \supp(\sigma)$, which ensures that the relative entropy is finite. Within this setting, we assume that $\rho$ is non-invertible, while $\sigma$ may or may not be invertible. This covers all cases not addressed by Petz’s original analysis in \cite{Petz:2003}.

To establish the monotonicity of relative entropy in this broader context, we must first ensure that $S(\Tr_b(\rho)\|\Tr_b(\sigma)) < +\infty$, i.e. that $\supp(\Tr_b(\rho)) \subseteq \supp(\Tr_b(\sigma))$. Observe that it is not necessary to check this condition when $\rho,\sigma > 0$, since in that case $\Tr_b(\rho)$ and $\Tr_b(\sigma)$ are positive-definite as well, and thus their supports coincide with $\h_a$.

We need a preliminary lemma regarding the kernel of positive semi-definite operators.

\blem\label{Lemma:ker}
	Let $T_1$ and $T_2$ be Hermitian positive semi-definite operators on a finite-dimensional Hilbert space $\h$. Then
	\beq
	\ker(T_1 + T_2) = \ker(T_1) \cap \ker(T_2).
	\eeq    
\elem

\proof
	Proving the inclusion $\ker(T_1) \cap \ker(T_2)\subseteq \ker(T_1 + T_2)$ is trivial: if we have $x \in \ker(T_1) \cap \ker(T_2)$, then $0_\h = T_1x + T_2x = (T_1 + T_2)x$, so $x \in \ker(T_1 + T_2)$.
	
	To show the inclusion $\ker(T_1 + T_2) \subseteq \ker(T_1) \cap \ker(T_2)$, we first notice that, thanks to the positive-semidefiniteness of $T_1$ and $T_2$, for all $x\in \h$ we have 
	\beq
	\langle (T_1 + T_2)x, x \rangle = \langle T_1x, x \rangle + \langle T_2x, x \rangle \geq 0,
	\eeq
	with equality to 0 if and only if $\langle T_1x, x \rangle = \langle T_2x, x \rangle = 0$, i.e. if and only if $T_1x=T_2x=0$ because the eigenvalues of both $T_1$ and $T_2$ are non-negative. 
	
	Hence, if $x \in \ker(T_1 + T_2)$, then
	$\langle (T_1 + T_2)x, x \rangle = 0$, so, due to the previous considerations, $T_1x=T_2x=0$, and therefore $x\in \ker(T_1) \cap \ker(T_2)$.
\qed

\bprop
	For all density operators $\rho,\sigma$ such that $\operatorname{supp}(\rho)\subseteq \operatorname{supp}(\sigma)$, it holds that
	\beq\label{eq:supptrb}
	\operatorname{supp}(\operatorname{Tr}_b(\rho)) \subseteq \operatorname{supp}(\operatorname{Tr}_b(\sigma)). 
	\eeq
\eprop

\proof
	Using the notation of section \ref{subsec:relentropy}, consider again the spectral decompositions of $\rho$ and $\sigma$
	\beq
	\rho=\sum_{j\in I_+(\rho)}\lambda_j P_j+0P_0, \quad \sigma=\sum_{k\in I_+(\sigma)}\mu_k \Pi_k + 0\Pi_0,
	\eeq
	where $P_j$ and $\Pi_k$ are the orthogonal projectors onto the eigenspaces corresponding to the positive eigenvalues $\lambda_j$ of $\rho$ and $\mu_k$ of $\sigma$, respectively. Then, the following operator sums yield the orthogonal projectors onto $\supp(\rho)$ and $\supp(\sigma)$, respectively
	\beq\label{eqconditiononsupports}
	P=\sum_{j\in I_+(\rho)}P_j, \quad \Pi=\sum_{k\in I_+(\sigma)}\Pi_k.
	\eeq
	Using the linearity and positivity of the partial trace and the fact that multiplying a positive semi-definite operator by a strictly positive scalar does not change its support, we have
	\beq 
	\begin{split}
		\supp\left(\Tr_b(\rho)\right) & = \supp\left(\sum_{j \in I_+(\rho)} \lambda_j \Tr_b(P_j)\right) = \supp\left(\sum_{j \in I_+(\rho)} \Tr_b(P_j)\right)\\
		& = \supp\left(\Tr_b \left(\sum_{j \in I_+(\rho)} P_j\right)\right) = \supp(\Tr_b(P)).
	\end{split}
	\eeq 
	The same argument applies to $\sigma$ and $\Pi$, so that
	\beq 
	\supp(\Tr_b(\sigma)) = \supp(\Tr_b(\Pi)).
	\eeq 
	Moreover, according to a standard result about orthogonal projectors, the inclusion $\supp(\rho)\subseteq\supp(\sigma)$ is equivalent to the fact that the operator $Q:=\Pi-P$ is an orthogonal projector on $\supp(\sigma)\cap \supp(\rho)^\perp=\supp(\sigma)\cap \ker(\rho)$. By the linearity of the partial trace, we can write
	\beq\label{traceprojectorinequality}
	\Tr_b(\Pi)=\Tr_b(P)+\Tr_b(Q),
	\eeq
	moreover, using the fact that $\Tr_b$ is a positive map and Lemma \ref{Lemma:ker}, we have:
	\beq
	\begin{split}
		\supp(\Tr_b(\Pi))&=\supp(\Tr_b(P)+\Tr_b(Q))\\
		&=\left[\ker(\Tr_b(P)+\Tr_b(Q))\right]^\perp\\
		&=\left[\ker(\Tr_b(P))\cap\ker(\Tr_b(Q))\right]^\perp\\
		&=\operatorname{span}(\ker(\Tr_b(P))^\perp \cup \ker(\Tr_b(Q))^\perp)\\
		& =\supp(\Tr_b(P))+\supp(\Tr_b(Q))\supseteq\supp(\Tr_b(P)),
	\end{split}
	\eeq
	having used a standard property of the orthogonal complement of the intersection of two vector subspaces. Therefore, 
	\beq
	\supp(\Tr_b(\rho))=\supp(\Tr_b(P))\subseteq\supp(\Tr_b(\Pi))=\supp(\Tr_b(\sigma)),
	\eeq
	as claimed.
\qed

While the support-based definition of relative entropy is useful to prove eq. \eqref{eq:supptrb}, in order to prove the monotonicity of relative entropy for a non-invertible density operator $\rho$, the regularized definition \eqref{epsilondefofRQE} turns out to be more adequate.

Since both $\rho_\varepsilon > 0$ and $\sigma_\varepsilon > 0$ for any arbitrarily small $\varepsilon > 0$, the modular operator $\Delta_{\rho_\varepsilon,\sigma_\varepsilon}$ is well-defined. Similarly, the modular operator $\Delta^a_{\rho_\varepsilon,\sigma_\varepsilon}$ is also well-defined. 

By applying the same steps used in the corrected version of Petz’s proof and using equation \eqref{eq:Srhost}, we obtain that, for any $\varepsilon > 0$
\beq
\begin{split}
	\Tr[\Tr_b(\rho_\varepsilon)\log\Tr_b(\rho_\varepsilon)-\Tr_b(\rho_\varepsilon)\log \Tr_b(\sigma_\varepsilon)] & =\langle \Tr_b(\rho_\varepsilon)^{1/2}, -\log(\Delta_{\rho_\varepsilon,\sigma_\varepsilon}^a)\Tr_b(\rho_\varepsilon)^{1/2}\rangle_a\\
	&\leq \langle\rho_\varepsilon^{1/2}, -\log(\Delta_{\rho_\varepsilon,\sigma_{\varepsilon}})\rho_{\varepsilon}^{1/2}\rangle_{ab}\\
	& = \Tr[\rho_\varepsilon\log\rho_\varepsilon-\rho_\varepsilon\log\sigma_\varepsilon].
\end{split}
\eeq
Notice that we can deal with $\rho_\varepsilon,\sigma_\varepsilon$ instead of $\rho,\sigma$ even if they do not have unit trace because this property is used in Petz’s original (and flawed) argument only in equation \eqref{eq:traceone}, which does not play any role in the corrected version outlined in Section \ref{susbsec:Petzcorrection}.

By continuity, taking the limit $\varepsilon\to 0^+$ on both sides of the previous inequality leads to
\beq
\begin{split}
	S(\Tr_b(\rho)||\Tr_b(\sigma)) & =\lim_{\epsilon\to 0^+}\Tr[\Tr_b(\rho_\varepsilon)\log\Tr_b(\rho_\varepsilon)-\Tr_b(\rho_\varepsilon)\log\Tr_b(\sigma_\varepsilon)]\\
	&\quad\leq\lim_{\epsilon\to 0^+}\Tr[\rho_\varepsilon\log\rho_\varepsilon-\rho_\varepsilon\log\sigma_\varepsilon]=S(\rho||\sigma),
\end{split}
\eeq
which completes the proof for non-invertible $\rho$ and general $\sigma$.

As a final remark, we note that, in \cite{Petz:2003}, Petz claimed that his proof applies not only to quantum channels but also to adjoints of unital Schwarz maps. However, that claim relies on the flawed argument that we have pointed out. We have also shown that the validity of Petz's proof can be restored by proving that $V_\rho$ is an isometry, a property that holds if we are dealing with  the partial trace but not with adjoint of a general unital Schwarz map.

\section{Uhlmann's proof of the monotonicity of the relative entropy under partial trace}\label{sec:Uhlmannproof}

The proof offered by Uhlmann in \cite{uhlmann1977relative} (see also \cite{Perez:2023} for a review) was more general than that of Petz because it naturally encompassed also the case of non-invertible density operators. However, thanks to the correction and generalization of Petz's proof outlined in the previous section, now we can state the two procedures have the same generality. 

Uhlmann's proof is based on the concept of interpolations of positive sesquilinear forms, that we recall in the following subsection. Coherently with the analysis developed so-far, we will consider only the case of finite dimensional vector spaces. 

\subsection{Interpolations of positive sesquilinear forms}\label{subsec:interpol}

Let $V$ be a vector space of finite dimension $d$ over the field $\mathbb{F}=\mathbb{R}$ or $\mathbb{C}$, and let $\mathcal{F}(V)$ be the set of sesquilinear forms\footnote{The results of this subsection encompass also the case of \textit{bilinear} forms.} over $V$, assumed to be linear in the second variable and conjugate-linear in the first. 

We say that $\alpha\in \mathcal{F}(V)$ is \textit{positive} if $\alpha(v,v)\geq 0$ $\forall v\in V$ and we denote the space of positive sesquilinear forms over $V$ by $\mathcal{F}_+(V)$. 

We can endow $\mathcal{F}_+(V)$ with a Löwner-like partial ordering: given $\alpha,\beta\in \mathcal{F}_+(V)$, we say that $\beta\le \alpha$ if $\alpha-\beta\geq 0$.

Fixing a basis of $V$, for any form $\alpha\in \mathcal{F}(V)$ there exists a unique Hermitian operator $T\in \text{End}_{\mathbb F}(V)$ such that for all $v, w \in V$, written in coordinates as $x, y \in \mathbb{F}^d$, one has
\beq\label{eq:opM}
\alpha(v, w) = x^\dagger T y.
\eeq 
It can be easily proven that, for a positive form $\alpha \in \mathcal{F}_+(V)$, the kernel of $\alpha$ coincides with the \textit{isotropic cone}
\beq 
\ker(\alpha) = \{v \in V \,:\, \alpha(v,v) = 0\},
\eeq
and they are both equal to the kernel of the positive semi-definite operator $T$.

Let now $\h$ be a Hilbert space with inner product $\langle \, , \, \rangle_\h$, $h \colon V \to \h$ be a linear surjective map and $A\in \bound_H(\h)$. Then $\alpha\in \mathcal{F}_+(V)$ is said to be \textit{represented by} $(\h,h,A)$, indicated with $\alpha\sim (\h,h, A)$, if
\beq
\alpha(v,w) = \langle h(v), A\, h(w)\rangle_\h\, ,\qquad \forall v,w\in V.
\eeq
Two representations of positive sesquilinear forms $\alpha\sim (\h,h,A)$ and $\beta\sim (\h,h,B)$ are said to be \textit{compatible} if $[A,B]=0$. As we shall see shortly, compatibility is a key concept in constructing a functional calculus for sesquilinear forms.

The following theorem shows that compatible representations exist and its constructive proof provides a (non-unique) way to build them.

\bt\label{compatible-th}
	Let $\alpha,\beta\in \mathcal{F}_+(V)$. Then there exist representations $\alpha\sim (\h,h,A)$ and $\beta\sim (\h,h,B)$ (with the same mapping $h$) such that  $[A,B]=0$.
\et

\proof
	Let $N\subset V$ be the kernel of the form $\alpha + \beta$
	\beq
	N=\{v\in V \, :\, \alpha(v,v)+\beta(v,v)=0\}.
	\eeq
	By fixing a basis of $V$, we can associate to $\alpha$ and $\beta$ two Hermitian and positive semi-definite operators $T_1,T_2\in \text{End}_{\mathbb F}(V)$ via eq. \eqref{eq:opM}. It follows that 
	\beq\label{eq:Nker}
	N=\ker(\alpha+\beta) = \ker(T_1+T_2)=\ker(T_1)\cap \ker(T_2),
	\eeq 
	where the last equality is provided by Lemma \ref{Lemma:ker}.
	
	Now, by setting $\h:=V/N$ we can define the following inner product:
	\begin{equation}
		\begin{array}{cccl}
			\langle \, , \, \rangle : &  \h\times\h & \longrightarrow &  \mathbb{F} \\
			&  (v+N,w+N) & \longmapsto         & \langle v+N,w+N\rangle = \alpha(v,w)+\beta(v,w).
		\end{array}
	\end{equation}
	Note that this inner product is well-defined thanks to the equalities \eqref{eq:Nker}.
	
	By taking $h$ as the (surjective) quotient map $h:V\to \h$, $v\mapsto h(v)=v+N$, we can define the following positive sesquilinear forms on $\h$:
	\beq
	\tilde \alpha(h(v),h(w))=\tilde \alpha(v+N,w+N):=\alpha(v,w),
	\eeq
	\beq
	\tilde \beta(h(v),h(w))=\tilde \beta(v+N,w+N):=\beta(v,w),
	\eeq
	for all $v,w\in V$.
	
	Thanks to the Riesz representation theorem, there exists a unique couple of positive Hermitian operators $A,B\in \bound_H(\h)$ such that 
	\beq
	\tilde \alpha(h(v),h(w))= \langle h(v), A \, h(w)\rangle = \alpha(v,w),
	\eeq
	\beq 
	\tilde \beta(h(v),h(w))= \langle h(v), B \, h(w)\rangle = \beta(v,w).
	\eeq
	It follows that, for all $v,w\in V$:
	\beq
	\begin{split}
		\langle h(v), h(w)\rangle &= \alpha(v,w)+\beta(v,w)=\tilde \alpha(h(v),h(w))+\tilde \beta(h(v),h(w)) \\
		& = \langle h(v), (A+B) \, h(w)\rangle,
	\end{split}
	\eeq 
	hence $A+B=id_\h$, which implies that $A$ and $B$ commute. 
\qed 

Consider now $\mathbb{R}^2_+=\{(x,y)\in  \mathbb{R}^2\, : \,  x\geq 0,\, y\geq 0\}$ and let $J$ be the set of homogeneous (of degree 1), measurable and locally bounded functions $f: \mathbb{R}^2_+\to \mathbb{R}$. It is possible to develop the concept of function of positive sesquilinear forms thanks to the following theorem, whose quite lengthy proof can be consulted in \cite{pusz1975functional}. 
\bt\label{indep-th}
	Let $V$ be a vector space, $\alpha,\beta\in \mathcal{F}_+(V)$ and let $\alpha\sim (\h,h,A)$ and $\beta\sim (\h,h,B)$ be two compatible representations of $\alpha$ and $\beta$. Then, for any $f\in J$, the function $\gamma:V\times V\to \mathbb F$, 
	\beq\label{explicitfunctionalofsesquilinearforms}
	\gamma(v,w):=\langle h(v), f(A,B) h(w)\rangle,  \qquad v,w\in V,
	\eeq
	is a well-defined sesquilinear form on $V$, i.e. $\gamma $ is independent of the choice of the representations and, for any given $f$, $\gamma$ depends only on $\alpha,\beta\in \mathcal{F}_+(V)$.
\et
Note that on the right-hand side of \eqref{explicitfunctionalofsesquilinearforms}, $f(A,B)$ is intended as an operator function, which is well-defined since $A$ and $B$ commute and so they can be simultaneously diagonalized.

Combining Theorem \ref{compatible-th} and Theorem \ref{indep-th}, every $f\in J$ can be extended to a function of positive sesquilinear forms. Keeping the same symbol for simplicity, we can define
\beq
f: \mathcal{F}_+(V)\times\mathcal{F}_+(V)\longrightarrow \mathcal{F}(V),
\eeq
as follows: given $\alpha,\beta\in \mathcal{F}_+(V)$ and any compatible representations $\alpha\sim (\h,h,A)$ and $\beta\sim (\h,h,B)$, the sequilinear form $f(\alpha,\beta)$ is represented as 
\beq\label{functionrepresentation}
f(\alpha,\beta)\sim(\h,h, f(A,B)).
\eeq
The elements of $J$ used to define the concept of interpolation of positive sesquilinear forms are the positive functions $f^t(x,y)= x^{1-t}y^t$, where $t\in[0,1]$. Given $\alpha,\beta\in \mathcal{F}_+(V)$, we call \textit{interpolation from $\alpha $ to $\beta$} the positive sesquilinear form 
\begin{equation}
	\begin{array}{cccl}
		\gamma^t_{\alpha\to \beta}:= f^t(\alpha,\beta) : &  V \times V  & \longrightarrow &  \mathbb{F} \\
		&  (v,w) & \longmapsto & f^t(\alpha,\beta)(v,w),
	\end{array}
\end{equation}
which means that for any compatible representations $\alpha\sim (\h,h,A)$ and $\beta\sim (\h,h,B)$,
\beq\label{explicitinterpolationform}
\begin{split}
	\gamma^t_{\alpha\to \beta}(v,w)&= \langle h(v), A^{1-t}B^t \hspace{1mm} h(w)\rangle,\qquad \forall v,w\in V.
\end{split}
\eeq
The interpolation is said to go from $\alpha$ to $\beta$ because, clearly, $\gamma^0_{\alpha\to \beta}=\alpha$ and $\gamma^{1}_{\alpha\to \beta}=\beta$. Moreover, the interpolation of two interpolations is another interpolation:  given $t_1,t_2\in [0,1]$, we have
\beq\label{interpolationofinterpolation}
\begin{split}
	\gamma^t_{\gamma^{t_1}_{\alpha\to \beta}\to \gamma^{t_2}_{\alpha\to \beta}}(v,w)&=\langle h(v) , (A^{1-t_1}B^{t_1})^{1-t}(A^{1-t_2}B^{t_2})^t \, h(w)\rangle\\
	&=\langle h(v) , A^{1-t_1-t+t_1t}A^{t-t_2t}B^{t_1-t_1t}B^{t_2t} \, h(w)\rangle\\
	&=\langle h(v) , A^{1-(t_1(1-t)+t_2t)}B^{t_1(1-t)+t_2t} \, h(w)\rangle\\
	&= \gamma^{t'}_{\alpha\to \beta}(v,w),
\end{split}
\eeq
with $t'=t_1(1-t)+t_2t$.

The interpolation of $\alpha,\beta$ computed at the value $t=1/2$ corresponds to a particularly important positive sesquilinear form, indicated with 
\beq
\sqrt{\alpha\beta}:=\gamma^{1/2}_{\alpha\to \beta},
\eeq 
and called \textit{geometric mean} of $\alpha,\beta$. Clearly, we have
\beq
\sqrt{\alpha\beta}\,(v,w)=\gamma^{1/2}_{\alpha\to \beta}(v,w) = \langle h(v),A^{1/2}B^{1/2}\, h(w)\rangle, \qquad \forall v,w\in V.
\eeq 
An important property of the geometric mean of $\alpha$ and $\beta$ will emerge in connection with the following concept: 
a positive sesquilinear form $r\in \mathcal{F}_+(V)$ is said to be \textit{dominated} by $\alpha,\beta\in \mathcal{F}_+(V)$ if
\beq
|r(v,w)|^2\leq \alpha(v,v)\beta(w,w), \qquad \forall v,w\in V.
\eeq
The following theorem establishes that $\sqrt{\alpha \beta}$ is the `maximal' positive sesquilinear form dominated by $\alpha$ and $\beta$. 

\bt\label{geomeancharacterization}
	Let $V$ be a vector space and let $\alpha,\beta \in \mathcal{F}_+(V)$, then $\sqrt{\alpha\beta}$ is dominated by $\alpha ,\beta$. Moreover, interpreting $\mathcal{F}_+(V)$ as a partially ordered set, let $S\subset \mathcal{F}_+(V)$ be the subset of positive sesquilinear forms dominated by $\alpha,\beta$, then $\sqrt{\alpha \beta}=\sup S$, i.e.  $r\leq \sqrt{\alpha\beta}$ for all $r\in S$.
\et
\proof
	The first statement is simply an application of the Cauchy-Schwarz inequality. For all $ v,w\in V$ and for any compatible representations of $\alpha$ and $\beta$ we have:
	\beq
	\begin{split}
		|\langle h(v), A^{1/2}B^{1/2} h(w)\rangle|^2&= |\langle A^{1/2} h(v),B^{1/2} h(w)\rangle|^2 \\
		&\leq \langle A^{1/2} h(v),A^{1/2} h(v)\rangle \langle B^{1/2} h(w),B^{1/2} h(w)\rangle\\
		&=\langle h(v),A h(v)\rangle\langle h(w),B h(w)\rangle,
	\end{split}        
	\eeq
	having used the fact that $A$ and $B$, and so their square roots, are Hermitian operators. 
	Thus, 
	\beq 
	|\sqrt{\alpha\beta}(v,w)|^2\leq  \alpha(v,v)\beta(w,w), \qquad \forall v,w\in V,
	\eeq
	so $\sqrt{\alpha\beta}$ is dominated by $\alpha$ and $\beta$.
	
	To prove the second statement, consider a form $r\in \mathcal{F}_+(V)$ dominated by $\alpha,\beta$ and let $\alpha\sim (\h,h,A)$ and $\beta\sim(\h,h,B)$ be compatible representations. By applying the constructive proof of Theorem \ref{compatible-th} to $r$, we can find a positive operator $C\in \bound_H(\h)$, such that  $r(v,w)=\langle h(v), C h(w)\rangle$, however the representation $r\sim (\h,h,C)$ will not, in general, be compatible with those of $\alpha$ and $\beta$. Since $r$ is dominated by $\alpha$ and $\beta$, we have
	\beq \label{dominatedrepr}
	|\langle h(v), C h(w)\rangle|^2\leq \langle h(v),A h(v)\rangle\langle h(w),B h(w)\rangle, \qquad \forall v,w\in V,
	\eeq
	or, thanks to the fact that $h$ is surjective on $\h$, 
	\beq\label{equivalentofscondstatement}
	|\langle x, C y\rangle|^2\leq \langle x,A x\rangle \langle y,B y\rangle, \qquad \forall x,y\in \h.
	\eeq
	Now, the second statement of the theorem, i.e. $r\le \sqrt{\alpha \beta}$, means that $(\sqrt{\alpha \beta}-r)(v,v)\ge 0$ for all $v\in V$, and all $r\in S$, which is equivalent to $\langle u,(A^{1/2}B^{1/2}-C)u\rangle \ge 0$ for all $u\in \h$ and all $C$ satisfying inequality \eqref{equivalentofscondstatement}. 
	
	It follows that the second statement of the theorem will be proven if we manage to show that $C\le A^{1/2}B^{1/2}$. In order to obtain this result, a regularization procedure applied on the operators $A,B$ is helpful: since $A\geq 0$ and $B\geq 0$, for all $\varepsilon>0$, $A_\varepsilon:=A+\varepsilon id_\h$, $B_\varepsilon:= B +\varepsilon id_\h$ and their square roots $A_\varepsilon^{1/2}$, $B_\varepsilon^{1/2}$ are Hermitian, positive and invertible. Moreover, it is clear that $A\leq A_\epsilon$, $B\leq B_\epsilon$, hence
	\beq\label{eq:ineqs}
	A_\varepsilon^{-1/2} A A_\varepsilon^{-1/2} \le id_\h, \quad B_\varepsilon^{-1/2} B B_\varepsilon^{-1/2} \le id_\h.
	\eeq 
	Finally, $A_\varepsilon$ and $B_\varepsilon$ are monotonically decreasing in $\varepsilon$ w.r.t. the Löwner ordering and  $A_\varepsilon \to A$ and $B_\varepsilon \to B$, as $\varepsilon \to 0$. For all $x,y\in \h$, there exist unique vectors $u,v\in \h$ such that
	\beq 
	x := A_\varepsilon^{-1/2} u, \quad y := B_\varepsilon^{-1/2} v,
	\eeq 
	then, inequality \eqref{equivalentofscondstatement} becomes
	\beq
	|\langle A_\varepsilon^{-1/2} u, C B_\varepsilon^{-1/2} v \rangle|^2 \leq \langle A_\varepsilon^{-1/2} u, A A_\varepsilon^{-1/2} u \rangle  \langle B_\varepsilon^{-1/2} v, B B_\varepsilon^{-1/2} v \rangle,
	\eeq 
	i.e.
	\beq
	|\langle  u, A_\varepsilon^{-1/2} C B_\varepsilon^{-1/2} v \rangle|^2 \leq \langle u, A_\varepsilon^{-1/2} A A_\varepsilon^{-1/2} u \rangle  \langle v, B_\varepsilon^{-1/2} B B_\varepsilon^{1/2} v \rangle\le \langle u,u\rangle \langle v,v\rangle,
	\eeq 
	having used the inequalities written in \eqref{eq:ineqs}. 
	
	By considering $u=v$, and taking into account that $A_\varepsilon^{-1/2}C  B_\varepsilon^{-1/2}$ is a positive operator, we can write
	\beq
	\langle u, A_\varepsilon^{-1/2}C  B_\varepsilon^{-1/2}u\rangle\le \langle u,u\rangle, \qquad \forall u\in \h,
	\eeq
	which implies that $A_\epsilon^{-1/2}C B_\epsilon^{-1/2}\leq id_\h$, thus $C\leq A_\varepsilon^{1/2} B_\varepsilon^{1/2}$. By taking the limit $\varepsilon\to 0$, we get $C\le A^{1/2}B^{1/2}$ and so also the second statement of the theorem is proven.
\qed

The property of the geometric mean just proven allows us to extend the ordering relation between two positive sesquilinear forms to their interpolations, in the sense specified by the following theorem.

\bt\label{interpolationineq}
	Let $V$ be a vector space and let $\alpha, \alpha', \beta,\beta'\in \mathcal{F}_+(V)$ such that  $\alpha'\le \alpha$ and $\beta'\le \beta$, then
	\beq
	\gamma^t_{\alpha'\to \beta'}\leq \gamma^t_{\alpha\to \beta}, \qquad \forall t\in [0,1].
	\eeq
\et

\proof
	The statement is clearly satisfied for $t=0,1$. Setting $t=1/2$, thanks to the first part of Theorem \ref{geomeancharacterization}, we get 
	\beq
	|\gamma^{1/2}_{\alpha' \to \beta'}(v,w)|^2\leq \alpha'(v,v)\beta'(w,w)\leq\alpha(v,v)\beta(w,w), \qquad \forall v,w\in V,
	\eeq
	where the second inequality follows from the hypotheses of this theorem. This means that $\gamma^{1/2}_{\alpha'\to \beta'}$ is dominated by $\alpha,\beta$ and so the extremality of the geometric mean implies:
	\beq\label{ineqgeommeans}
	\gamma^{1/2}_{\alpha'\to \beta'}\leq \gamma^{1/2}_{\alpha\to \beta}.
	\eeq
	If we now use eq. $\eqref{interpolationofinterpolation}$ with $t=t_2=1/2$ and $t_1=0$, we get
	\beq
	\gamma^{1/4}_{\alpha\to \beta}=\gamma^{1/2}_{\gamma^0_{\alpha\to \beta}\rightarrow \gamma^{1/2}_{\alpha\to \beta}},\qquad \gamma^{1/4}_{\alpha'\to \beta'}=\gamma^{1/2}_{\gamma^0_{\alpha'\to \beta'}\rightarrow \gamma^{1/2}_{\alpha'\to \beta'}}.
	\eeq
	By repeating the previous argument, this time using eq. \eqref{ineqgeommeans}, we show that $\gamma^{1/4}_{\alpha'\to \beta'}\leq \gamma^{1/4}_{\alpha\to \beta}$.
	
	By iterating this procedure, we can prove the statement of the theorem for any $t\in [0,1]$ of the type $t_{k,n}=k/2^n$, with $n,k\in \mathbb{N}$, $k\leq 2^n$, which is a dense subset of $[0,1]$. 
	
	Finally, the functions $t_{k,n}\to \gamma^{t_{k,n}}_{\alpha\to \beta}(v,w)$ and $t_{k,n}\to \gamma^{t_{k,n}}_{\alpha'\to \beta'}(v,w)$ are continuous for every fixed $v,w\in V$, hence the theorem holds for all $t\in [0,1]$.
\qed

In a similar way, we can prove another important result. Let $\psi:U\to V$ be a linear map between vector spaces and let $\alpha,\beta\in \mathcal{F}_+(V)$. Then, $\psi$ allows us to \textit{pull-back} these sesquilinear forms on $U$ as follows
\begin{equation}
	\begin{array}{cccl}
		\psi^*\alpha : &  U\times U & \longrightarrow &  \mathbb{F} \\
		&  (v,w) & \longmapsto         & \alpha(\psi(v),\psi(w))
	\end{array},\qquad
	\begin{array}{cccl}
		\psi^*\beta : &  U\times U & \longrightarrow &  \mathbb{F} \\
		&  (v,w) & \longmapsto         & \beta(\psi(v),\psi(w))
	\end{array}.
\end{equation}

The following theorem shows that the pull-back of an interpolation of forms in $\mathcal F_+(V)$ is always `smaller' than the interpolation of their pull-backs, with respect to the partial ordering of $\mathcal F_+(U)$.

\bt\label{pull-backineq}
	Let $U,V$ be vector spaces, $\psi:U\to V$ be a linear map and $\alpha,\beta\in \mathcal{F}_+(V)$, then 
	\beq\label{eq:inegpullback}
	\psi^*\gamma^t_{\alpha\to \beta}\leq\gamma^t_{\psi^*\alpha\to \psi^*\beta},\qquad\forall t\in[0,1].
	\eeq
\et
\proof
	The argument that we use is quite similar to the one appearing in the previous proof. The statement is true for $t=0$ and $t=1$ because in these cases we have $\psi^*\alpha\le \psi^*\alpha$ and $\psi^*\beta\le \psi^*\beta$, respectively. 
	
	Let us now consider $t=1/2$, which gives rise to the geometric mean, so
	\beq
	|\psi^*\sqrt{\alpha\beta}(v,w)|^2= |\sqrt{\alpha\beta}(\psi(v),\psi(w))|^2, \qquad \forall v,w\in U.
	\eeq 
	Since $\sqrt{\alpha\beta}$ is dominated by $\alpha,\beta$, we can write
	\beq
	|\psi^*\sqrt{\alpha\beta}(v,w)|^2 
	\leq \alpha(\psi(v)\psi(v))\beta(\psi(w),\psi(w))
	=\psi^*\alpha(v,v)\psi^*\beta(w,w),
	\eeq
	which shows that $\psi^*\sqrt{\alpha\beta}$ is dominated by $\psi^*\alpha$, $\psi^*\beta$. Thus, by the extremal property of the geometric mean, the statement of the theorem holds for $t=1/2$. 
	
	By iterating this reasoning as done in the proof of the previous theorem, the validity of inequality \eqref{eq:inegpullback} can be generalized to all $t\in [0,1]$. 
\qed

\subsection{Definition of the relative entropy in terms of interpolations of forms}

In this subsection, we apply the results previously established to reformulate the relative entropy in a manner that will facilitate the proof of its monotonicity under partial trace, to be presented in the next subsection.

Adopting analogous notations as those introduced at the beginning of Section \ref{sec:Petzproof}, we identify the vector space $V$, on which the forms of interest for us will be defined, with $\bound(\h_{ab})$. As we know, this is a Hilbert space w.r.t. the Hilbert-Schmidt inner product $\langle A,B\rangle_{ab} = \Tr(A^\dagger B)$, $A,B\in \bound(\h_{ab})$,  which is a positive-definite sesquilinear form. 

Given two density operators $\rho,\sigma\in \bound_H(\h_{ab})$, we can define the following positive sesquilinear forms $\rho_L,\sigma_R : \bound(\h_{ab})\times\bound(\h_{ab}) \to  \mathbb{F}$,
\begin{equation}\label{eq:rhoL}
	\rho_L(A,B):=\Tr(\rho B A^\dag)= \Tr(A^\dag \rho B)= \langle A, L_{\rho}B\rangle,
\end{equation}
\begin{equation}\label{eq:sigmaR}
	\sigma_{R}(A,B):= \Tr(\sigma A^\dag B)=\Tr(A^\dag B \sigma)= \langle A, R_{\sigma}B\rangle,
\end{equation}
where the operators $L_\rho$ and $R_\sigma$ are defined as in eq. \eqref{right,left actions}. We immediately recognize the two representations 
\beq\label{representations}
\rho_{L}\sim (\bound(\h_{ab}), id_{ab}, L_{\rho}), \quad 
\sigma_{R}\sim (\bound(\h_{ab}), id_{ab}, R_{\sigma}),
\eeq
where $id_{ab}$ is the identity map on $\bound(\h_{ab})$. These representations are compatible because, thanks to eq. \eqref{eq:brackL}, $[L_{\rho},R_{\sigma}]=0$.

We can now define the \textit{relative entropy positive sesquilinear form} between the two density operators $\rho,\sigma$, indicated with $S_{\rho \| \sigma}:\bound(\h_{ab})\times\bound(\h_{ab}) \to  \mathbb{F}$, as the rate of change of the interpolation $\gamma^t_{\rho_{L}\to \sigma_{R}}$ with respect to $\rho_L$
\beq\label{interpolationversionofrel}
S_{\rho\|\sigma}(A,B)= -\liminf_{t\to 0^+}\frac{\gamma^t_{\rho_{L}\to \sigma_{R}}(A,B)- \rho_{L}(A,B)}{t},\qquad A,B\in \bound(\h_{ab}).
\eeq
\textit{Remark}. The use of $\liminf$ instead of an ordinary limit is motivated by the following two arguments:

\begin{enumerate}
	\item When $\rho$ and $\sigma$ are not invertible, the interpolation function
	$
	t \mapsto \gamma^t_{\rho_L\to  \sigma_R}(A, B)
	$
	may not be differentiable at $t = 0^+$. The ordinary limit of the differential quotient at $t=0^+$ might fail to exist due to oscillations or lack of smoothness in the interpolation path. In such situations, the ordinary limit does not exist. However, the $\liminf$ always exists (possibly infinite), thereby ensuring that the entropy form $S_{\rho \|\sigma}(A, B)$ is always well-defined.
	
	\item The function $t \mapsto \gamma^t_{\rho_L\to \sigma_R}(A, B)$ is convex in $t$, as it arises from the interpolation
	$
	f_t(x, y) = x^{1-t} y^t,
	$
	which is jointly operator convex for $x, y > 0$. For convex functions, the left and right derivatives at an endpoint may differ, and the correct notion of derivative from the right at $t = 0$ is the lower right Dini derivative, i.e. the $\liminf$ of the difference quotient. Thus, the use of $\liminf$ aligns with standard practice in convex analysis.
\end{enumerate}
The relative entropy between states $\rho,\sigma $ can be recovered as follows.

\bt For all density operators $\rho,\sigma \in \bound_H(\h_{ab})$ we have
	\beq\label{eq:redef}  
	S(\rho\|\sigma)=S_{\rho\|\sigma}(id_{ab},id_{ab}).
	\eeq
\et

\proof
	Let us first consider the case of invertible density operators $\rho>0$ and $\sigma>0$. Then, 
	\beq\label{proofRQMintermsofinterpolations}
	\begin{split}
		S_{\rho\|\sigma}(id_{ab},id_{ab})&=-\liminf_{t\to 0^+}\frac{\gamma^t_{\rho_{L}\to \sigma_{R}}(id_{ab},id_{ab})- \rho_{L}(id_{ab},id_{ab})}{t}\\
		&=-\liminf_{t\to 0^+}\frac{\langle id_{ab}, L_\rho^{1-t }R_\sigma^t id_{ab}\rangle-\langle id_{ab}, L_\rho id_{ab}\rangle}{t}\\
		&=-\liminf_{t\to 0^+}\frac{\Tr(\rho^{1-t}\sigma^t)-\Tr(\rho)}{t}=-\left. \frac{d}{dt}\right|_{t=0}\Tr\big(\rho^{1-t}\sigma^t)\\
		&=-\left. \frac{d}{dt}\right|_{t=0}\Tr\big(\exp((1-t)\log\rho)\exp(t\log\sigma)\big)\\
		&=-\Tr\big[-\exp((1-t)\log\rho)\log\rho\exp(t\log\sigma)\\
		&\qquad \quad \; \, +\exp((1-t)\log\rho)\exp(t\log\sigma)\log\sigma\big]_{t=0}\\
		&=-\Tr\big(-\rho\log\rho+\rho\log\sigma\big)=\Tr\big(\rho\log\rho-\rho\log\sigma\big)\\
		&=S(\rho\|\sigma).
	\end{split}
	\eeq
	In the case of $\sigma$ and $\rho$ that are not invertible, we can use the regularized version of relative entropy. We have seen an equivalent definition of the relative entropy in \eqref{epsilondefofRQE} and then it is natural to consider 
	\beq
	\lim_{\epsilon\to 0^+}  S_{\rho_\epsilon\|\sigma_\epsilon}(id_{ab}, id_{ab}).
	\eeq
	Now we have that both $\rho_\epsilon,\sigma_\epsilon$ are positive definite and, thus, $\log \rho_\epsilon$, $\log\sigma_\epsilon$ are well defined. By repeating the steps in \eqref{proofRQMintermsofinterpolations}, we get that 
	\beq
	\lim_{\epsilon\to 0^+ }S_{\rho_\epsilon\|\sigma_\epsilon}(id_{ab},id_{ab})=\lim_{\epsilon\to 0^+} \Tr(\rho_\epsilon \log\rho_\epsilon-\rho_\epsilon\log\sigma_\epsilon).
	\eeq
	Obtaining the same definition of relative entropy as in \eqref{RQEdef}.
\qed

\subsection{Proof of the monotonicity of the relative entropy under partial trace}

Thanks to formula \eqref{eq:redef}, the monotonicity of the relative entropy under the partial trace $\Tr_b:\bound(\h_{ab})\to \bound(\h_a)$ will be proven if we show that 
\beq
S_{\Tr_b(\rho)\|\Tr_b(\sigma)}(id_{a},id_{a})\le S_{\rho\|\sigma}(id_{ab}, id_{ab}).
\eeq
As in Petz’s proof, the adjoint of the partial trace plays a central, though conceptually distinct, role in establishing the monotonicity of relative entropy. 

Specifically, in Uhlmann’s approach, $\Tr_b^\dag$ acts as a pull-back map: we define $\psi := \Tr_b^\dag : \bound(\h_a) \to \bound(\h_{ab})$ and use it to pull back the positive sesquilinear forms $\rho_L$ and $\sigma_R$ introduced in Eqs. \eqref{eq:rhoL} and \eqref{eq:sigmaR}, respectively. Thanks to Theorem \ref{pull-backineq} we have
\beq
\psi^*\gamma^t_{\rho_L\to \sigma_R}\leq \gamma^t_{\psi^* \rho_L\to \psi^*\sigma_R},
\eeq
i.e.
\beq\label{firstpartineq}
\gamma^t_{\rho_L\to \sigma_R}(\Tr_b^\dag(X),\Tr_b^\dag(X))= \psi^*\gamma^t_{\rho_L\to \sigma_R}(X,X)\leq \gamma^t_{\psi^*\rho_L\to \psi^*\sigma_R}(X,X), 
\eeq
for all $X\in \bound(\h_a)$. Now we can use the fact that $\Tr_b$ preserves density operators and that $\Tr_b^\dag$ is a Schwarz map 
to write
\beq
\begin{split}
	\psi^*\rho_L (X,X)&=\rho_L(\Tr_b^\dag(X),\Tr_b^\dag(X))=\Tr(\Tr_b^\dag(X)^\dag\rho \Tr_b^\dag(X)) = \Tr(\rho \Tr_b^\dag(X)\Tr_b^\dag(X)^\dag) \\
	&\leq \Tr(\rho \Tr_b^\dag(XX^\dag))=\langle \rho,\Tr_b^\dag(XX^\dag)\rangle=\langle \Tr_b(\rho),XX^\dag\rangle = \Tr(\Tr_b(\rho)X X^\dag) \\
	& =\Tr_b(\rho)_L(X,X).
\end{split}
\eeq
Replacing $\rho_L$ with $\sigma_R$, we find
\beq
\psi^*\sigma_R(X,X)\leq \Tr_b(\sigma)_R(X,X).
\eeq
Thus, we have proven that $\psi^*\rho_L\le \Tr_b(\rho)_L$ and $\psi^*\sigma_R\le \Tr_b(\sigma)_R$ and so Theorem \ref{interpolationineq} implies that, for all $t\in [0,1]$, it holds that
\beq
\gamma^t_{\psi^*\rho_L \to\psi^*\sigma_R}\leq \gamma^t_{\Tr_b(\rho)_L
	\to \Tr_b(\sigma)_R}.
\eeq
This result and inequality \eqref{firstpartineq} imply
\beq
\gamma^t_{\rho_L\to \sigma_R}(\Tr_b^\dag(X),\Tr_b^\dag(X))\le \gamma^t_{\Tr_b(\rho)_L
	\to \Tr_b(\sigma)_R}(X,X),
\eeq 
for all $X\in \bound(\h_a)$. By considering in particular $X=id_a$ and recalling that $\Tr_b^\dag$ is a unital map, i.e. $\Tr_b^\dag(id_{a})=id_{ab}$, we obtain
\beq\label{eq:almostdone}
\begin{split}
	\gamma^t_{\rho_L \to\sigma_R}(id_{ab},id_{ab})\leq \gamma^t_{\Tr_b(\rho)_L\to \Tr_b(\sigma)_R}(id_a,id_a).
\end{split}
\eeq
Now, since $\Tr_b$ is trace-preserving, we have
\beq
\begin{split}
	\Tr_b(\rho)_L(id_a,id_a) & = \langle id_a, L_{\Tr_b(\rho)} id_a\rangle_a=\Tr(\Tr_b(\rho))=\Tr(\rho) = \langle id_{ab}, L_\rho id_{ab}\rangle_{ab}\\
	& =\rho_L(id_{ab},id_{ab}),
\end{split}
\eeq
thus, we can rewrite inequality \eqref{eq:almostdone} as follows
\beq\label{objineq}
\gamma^t_{\rho_L \to \sigma_R}(id_{ab},id_{ab})-\rho_L(id_{ab},id_{ab})\le\gamma^t_{\Tr_b(\rho)_L \to \Tr_b(\sigma)_R}(id_a,id_a)-\Tr_b(\rho)_L(id_a,id_a).
\eeq
Thanks to eqs. \eqref{eq:redef} and \eqref{interpolationversionofrel}, this last inequality implies the monotonicity of the relative entropy under partial trace.

\section{Conclusions}

We have revisited the monotonicity of relative entropy under the action of quantum channels by focusing on two important proofs: those by Petz and Uhlmann. While both approaches are foundational, their complexity has often hindered their pedagogical dissemination.

Our aim was to clarify and reconstruct these strategies within a finite-dimensional operator framework. In particular, we pointed out a subtle flaw in Petz's original argument, whose validity was nonetheless restored by Petz and Nielsen soon after, and showed how to rigorously extend this approach to incorporate non-invertible density operators.

It is also worth noting that our explicit construction of the isometric operator defined in equation \eqref{eq:defVs} sheds new light on its structural role within the broader context of quantum information theory. In particular, this operator can be seen as an essential component of the \textit{Petz recovery map}, originally introduced in \cite{Petz:88} in the setting of von Neumann algebras. Given a quantum channel $\mathcal{C}$ and a fixed full-rank state $\sigma \in \bound_H(\mathcal{H}_{ab})$, the Petz recovery map associated to $\mathcal{C}$ and $\sigma$ is defined by
\beq
\mathcal{P}_{\sigma, \mathcal{C}}(\rho) := \sigma^{1/2} \, \mathcal{C}^\dag\left( \mathcal{C}(\sigma)^{-1/2} \, \rho \, \mathcal{C}(\sigma)^{-1/2} \right) \, \sigma^{1/2},
\eeq
or equivalently, in terms of superoperators,
\beq
\mathcal{P}_{\sigma, \mathcal{C}} = L_{\sigma^{1/2}} \circ R_{\sigma^{1/2}} \circ \mathcal{C}^\dag \circ R^a_{\mathcal{C}(\sigma)^{-1/2}} \circ L^a_{\mathcal{C}(\sigma)^{-1/2}}.
\eeq
When $\mathcal{C} = \Tr_b$, this reduces to
\beq
\begin{split}
	\mathcal P_{\sigma,\Tr_b} &= L_{\sigma^{1/2}} \circ (R_{\sigma^{1/2}} \circ \mathcal \Tr_b^\dag \circ R^a_{\Tr_b(\sigma)^{-1/2}}) \circ L^a_{\Tr_b(\sigma)^{-1/2}}\\
	& = L_{\sigma^{1/2}} \circ V_\sigma \circ L^a_{\Tr_b(\sigma)^{-1/2}},
\end{split}
\eeq 
or
\beq
V_\sigma = L_{\sigma^{-1/2}} \circ \mathcal P_{\sigma,\Tr_b} \circ L^a_{\Tr_b(\sigma)^{1/2}}.
\eeq 
So, just as the Petz recovery map characterizes the reversibility of quantum channels and identifies conditions for saturation of the monotonicity inequality, the operator $V_\sigma$ captures explicitly the mechanism by which relative entropy is contracted under partial trace. 

In recent developments, particularly in the work of Fawzi and Renner \cite{FawziRenner:15}, the Petz recovery map plays a central role in quantitative refinements of the data processing inequality. Specifically, for states $\rho$, $\sigma$ and a channel $\mathcal{C}$, the inequality
\begin{equation}
	S(\rho\|\sigma) - S(\mathcal{C}(\rho)\|\mathcal{C}(\sigma)) \geq -2 \log F\left(\rho, (\mathcal{P}_{\sigma,\mathcal{C}} \circ \mathcal{C})(\rho)\right),
\end{equation}
bounds the loss of distinguishability in terms of the fidelity $F$ between the original state $\rho$ and its recovered approximation via $\mathcal{P}_{\sigma,\mathcal{C}} \circ \mathcal{C}$. 

The explicit identification of $V_\sigma$ offers a concrete realization of this recovery mechanism, reinforcing its interpretive clarity and suggesting further applications in entropy inequalities and recoverability conditions.

\section*{Acknowledgments}
The authors are grateful to Prof. Mark M. Wilde for his valuable insights on the literature concerning relative entropy and, in particular, for clarifying the correction of the flawed Petz argument through the proof that the operator $V_\rho$ is an isometry.

\bibliographystyle{plain} 
\bibliography{bibliography}

\end{document}